\documentclass[12pt,preprint]{aastex}
\shorttitle{Opposition Surge Cause}
\shortauthors{Schaefer, Rabinowitz, \& Tourtellotte}

\begin{document}

\title{The Diverse Solar Phase Curves of Distant Icy Bodies.  II. The Cause of the Opposition Surges and Their Correlations}

\author{Bradley E. Schaefer}
\affil{Physics and Astronomy, Louisiana State University,
    Baton Rouge, LA, 70803, USA; schaefer@lsu.edu}
\author{David L. Rabinowitz}
\affil{Center for Astronomy and Astrophysics, Yale University, New Haven, CT, USA}
\author{Suzanne W. Tourtellotte}
\affil{Department of Astronomy, Yale University, New Haven, CT, USA}

\begin{abstract}

We collect well-measured opposition surge properties for many icy bodies orbiting the Sun (mostly from our own observations) plus for many icy moons, resulting in a data base of surface and orbital properties for 52 icy bodies.  (1) We put forward four criteria for determining whether the surge is being dominated by shadow hiding (SH) or coherent backscattering (CB) based on readily measured quantities.  The CB surge mechanism dominates if the surge is color dependent, the phase curve is steeper than 0.04 mag deg$^{-1}$, the phase curve shape matches the CB model of Hapke, or if the albedo is higher than roughly 40\%.  (2) We find that virtually all of our sample have their phase curves dominated by CB at low phase angles.  (3) We present a graphical method to determine the Hapke surge parameters $B_{C0}$ and $h_C$.  (4) The Kuiper Belt Objects (KBOs) and Centaurs have relatively high surge amplitudes, $B_{C0} \gtrsim 0.5$ and widths with $h_C \sim 3\degr$.  (5) We find highly significant but loose correlations between surge properties and the colors, albedos, and inclinations.  We interpret this as young surfaces tending to have low surge slopes, high albedo, and gray colors.  (6) Nereid has its surface properties similar to other icy moons and greatly different from KBOs and Centaurs, so we conclude that Nereid is likely a nearly-ejected inner Neptunian moon rather than a captured KBO.

\end{abstract}

\section{Introduction}

	Planetary astronomers have few means to collect information about distant icy bodies in our Solar System.  The possibilities include photometry, spectroscopy, astrometry, and spacecraft visits; all have substantial limitations.  For photometry, almost all prior work has concentrated on obtaining colors and rotational light curves.  With the exception of a few bodies (notably Pluto, Charon, Nereid, Chiron, and Pholus), none of the small icy bodies in the outer Solar System have been observed for well-sampled phase curves or secular variations.  The reason is simply that these two questions require telescope time on many widely-spread nights and this is generally impossible within the normal mode of observing runs of a few nights per year.  

With the lack of any useful information on phase curves and secular variations, in 2003, our group decided to start a large long-term project to observe many small icy bodies on a large number of nights in three colors.  What made this possible was that the SMARTS telescopes on Cerro Tololo in Chile were being operated in a queue mode, wherein a resident observer would collect CCD images for many different programs all in one night with the programs continuing nightly for a year or more (Bailyn 2004).  With this unique observing mode, we had the opportunity to obtain photometry on every clear night, year-round, and year-after-year without the logistical difficulties of living at the observatory.  It has become possible to get nightly multi-color magnitudes for Kuiper Belt Objects (KBOs) for their entire apparition (spanning all possible phase angles) and for multiple apparitions (to check for secular variations).

	Our group started by using the SMARTS telescopes (then called YALO) in Chile to get extensive photometry of Nereid (Schaefer and Schaefer 2000; Schaefer and Tourtellotte 2001), and Huya (Ferrin et al. 2001; Schaefer and Rabinowitz 2002).  In 2003, with access to significant amounts of SMARTS telescope time, our group started the regular observations of many KBOs, cycling between targets as they passed through their apparitions.  The result is that we obtained very well-sampled ($\sim$100 nights per target) B-, V-, and I-band solar phase curves for 18 KBOs, 7 Centaurs, and Nereid (Rabinowitz, Schaefer, and Tourtellotte 2007, hereafter RST2007).  Our data base now constitutes nearly all of our community's observations of solar phase curves for KBOs and Centaurs.

	The main constraint of our program is that we are not using a large telescope, so we cannot get useful data for the fainter bodies.  Our primary telescope is the SMARTS 1.3 meter telescope with the ANDICAM CCD camera, which forces us to select targets brighter than roughly 21.5 magnitude (and preferrably much brighter).  We have already extended our program to look at Jupiter Trojan asteroids, and we will be further extending our program to cover more planetary moons and some main-belt asteroids that pass through near-zero phase angle.

	In this paper, we will use our large and unique data base from RST2007 to address two questions.  First, we will determine the physical mechanism responsible for the observed opposition surges.  The two possible causes are shadow hiding (SH) and coherent backscattering (CB).  We will establish four observational criteria for distinguishing these two causes.  Our result will be that apparently {\it all} our targets will have their opposition surges dominated by CB.  Second, we will correlate our observed opposition surge properties (along with other surface properties) with orbital properties.  We will consider a total of 52 icy bodies in the outer Solar System and group them by their size and orbital properties.  We find that the surge properties are significantly correlated with the color, albedo, and inclination of the bodies.
	
\section{The Data}

	For our study, we need a large and complete collection of objects with well-measured orbital, photometric, and surge properties.  Other than for planetary moons, almost all of the available surge data is from our own program (RST2007; Schaefer \& Rabinowitz 2002).  We supplement this with four new objects (120178=2003 OP32, 145453=2005 RR43, 24835=1995 SM55, and 145451=2005 RM43) for which our measures of their surge properties are presented in Rabinowitz et al. (2008, RSST2008).  For Nereid, we have additional yearly measures of the surge properties, colors, and variations (Schaefer et al. 2007).  Our own data base now consists of 30 sources for which we have very-well-sampled phase curves in three colors.
	
	For objects other than planetary moons, the literature contains many reports of sketchy phase curves, but unfortunately most are not usable for purposes of this paper.  Primary problems include that only two or three distinct phases are measured, the phase ranges are not adequate, and no rotational corrections are known.  From individual papers throughout the literature, we found only three objects with usable phase curves (with reasonable phase coverage and rotational corrections as needed even if only in one color); 40314=1999 KR16 (Sheppard \& Jewitt 2002), Echeclus (Rousselot et al. 2005), and Pelion (Bauer et al. 2003).  Another possible means of getting phase curve information from the literature is to piece them together from scattered papers (Belskaya et al. 2003).  To this end, we have collected a very large data base of reported photometry, with observations of individual bodies grouped together.  Primary problems are the presence of systematic offsets in some large data sources (a similar problem is reported by Tegler \& Romanishin 2003), the lack of adequate phase coverage, and the lack of rotational corrections as needed.  Out of all the many objects in our compilation, only two (Hylonome and Elatus) presented well-sampled and consistent phase curves, and these are two of the targets highlighted by Belskaya et al. (2003).  For the objects other than planetary moons, we have added only 5 sources based on data taken from the literature.
	
	For the planetary moons, we have taken from the literature the phase curve information for 15 moons of Jupiter, Saturn, Uranus, and Neptune.  The phase curve for Nereid is taken from Schaefer et al. (2007).  Pluto and Charon have their phase curves taken from Buie, Tholen, \& Wasserman (1997).
	
	Some bodies with adequate data had to be excluded for other reasons.  Chiron displays 'flares' on the time scale of years and shorter, such that no consistent phase curve can be distinguished from its secular variations.  The two new Plutonian moons, Nix and Hydra, do not yet have the data for their phase curves analyzed (Springman, et al. 2007; Buie et al. 2007).  The Saturnian and Uranian rings are small icy bodies in the outer Solar System for which there are excellent multi-color phase curves (French et al. 2007; Karkoschka 2001), but the physical conditions are so unique that they would each have to be placed alone in a group, and the addition of one new group to account for one target provides little understanding not in the original paper already.  Similarly, Io is not included as its surface properties are unique.  Titan is not included as its atmosphere hides the icy surface of interest.
	
	In all, we have collected phase curve information for 52 icy bodies in the outer Solar System.  We have also collected various other data for all 52 bodies, with these being collected into Table 1 (name, orbit, and size information), Table 2 (surface photometric data), and Table 3 (surge properties).  These 52 bodies can be categorized into a variety of classification schemes.  In Table 1, we will list the class for each body by standard nomenclature based on their sizes and orbits (see next paragraph).  However, in Tables 1-3, we will also collect the bodies into new classification groups based on their surface properties (see below).  A major point of this paper will be that the surface properties and groups are strongly correlated with the size and orbital properties. 
	
	Our 52 icy bodies have been placed into 7 classes based on their size and orbit.  The seven classes are Centaurs, Plutinos, classical KBOs, dwarf planets, the Haumea collisional group, scattered disk objects (SDOs) with aphelion distances (Q) greater than 70 AU, and planetary moons.  The Centaur population is bimodal in color (Tegler, Romanishin, \& Consolmagno 2003; Peixinho et al. 2003), so we have also identified for each object whether it is in the gray set or the red set.  The dwarf planet class is based on size as well as orbit, with the justification for separating them out from the classical KBOs and Plutinos being that their surface properties are expected to be different due to their size (Schaller \& Brown 2007).  The inclusion of other large bodies in this class is possible with Sedna, Charon, and Haumea being the obvious candidates.  The Haumea collisional class (formerly called the '2003 EL61 collisional group') is a set of several KBOs all in similar orbits with similar surface composition, which presumably was formed when some collision shattered fragments off the dwarf planet (Brown et al. 2007) roughly one billion years ago (Ragozzine \& Brown 2007).  Alternative classes are reasonable for some of the icy bodies, for example Sedna could be assigned as the only known member of an inner Oort Cloud group, while the Centaur and SDO populations intermix on short dynamical time scales so it might be reasonable to combine these into one group.
	
	Our 52 icy bodies have also been separated into 6 groups based on their surface photometric properties and size.  The names for these groups that we will adopt are Small/Red, Small/Gray, Intermediate, Large, Collisional, and Moons.  The Small/Red bodies are those Centaurs, KBOs, and SDOs smaller than ~800 km diameter and with B-R color redder than 1.5 mag.  The Small/Gray bodies are those Centaurs, KBOs, and SDOs smaller than ~800 km diameter and with B-R color grayer than 1.5 mag.  The Intermediate group is for those bodies with diameter from roughly 800-1400 km.  The Large group is for those icy bodies with diameters larger than roughly 1400 km.  The Collisional group is for those bodies with similar composition (and orbits) as Haumea.  The Moon group is for planetary moons.  Some bodies could belong to alternative groups, for example Charon and Ariel could both be placed into the Intermediate or Moon groups, with our resolution of these ambiguities based on the observed surface properties.  The goal in making these groups is to cluster bodies by physical mechanisms or histories that might dominate their surface properties.  
	
	In Table 1 we collect all the name and orbit data.  The first three columns give the proper name, number, and designation for each of our 52 icy bodies.  This listing can be used as a concordance between the alternative names.  The next three columns give orbital information including the inclination ($i$ in degrees), the eccentricity ($e$), and the semi-major axis ($a$ in AU).  These data are taken from the JPL Horizons web site\footnote{http://ssd.jpl.nasa.gov/horizons.cgi} or from the {\it Astronomical Almanac} (for the planetary orbit values).  The next two columns list two measures of the object's size.  The first measure is the absolute magnitude as tabulated from the JPL Horizons web site or from the {\it Astronomical Almanac}, but this value can only give an approximate measure of the size without knowing the albedo, so the absolute magnitude is only useful as a size indicator for bodies without an independently measured diameter.  The second measure gives the diameter of the body, as tabulated in Stansberry et al. (2007) or the {\it Astronomical Almanac}.  The last column gives the class identification for each object.
	
	In Table 2, we collect the surface information for our 52 icy bodies.  The first column lists the name, with the designation given if no name has been adopted.  The entries are in the same order as in Table 1, and the our new groups have been inserted as headers for each section.  The second column gives the geometric albedo ($p_V$) in the V-band as a percentage.  These albedos for the KBOs and Centaurs were all taken from Stansberry et al. (2007) as based on thermal infrared fluxes.  The albedos of the planetary moons were taken from the {\it Astronomical Almanac}, Karkoschka (1997) for the Uranian moons, and Buie, Tholen, \& Wassserman (1997) for Pluto and Charon.  We note that geometric albedos depend on the exact magnitude of the opposition surge, and many of the reported albedos are based on relatively poor phase curves or assumptions, where for example the presence of a narrow opposition 'spike' would substantially change $p_V$.  For purposes of this paper, we are only needing albedos as approximate indicators of surface reflectivity, so the $p_V$ values from the cited references are more than adequate.  An updating and systematization of all the geometric albedos is an important task that is beyond the scope of this paper.  The third and fourth columns list the B-I color (in magnitudes) evaluated at (or extrapolated to) zero phase angle along with the reference for that color.  The fifth column gives a short-hand summary of the surface composition as based on optical and infrared spectroscopy, with the citation presented in the last column.
	
	In Table 3, we collect together all available phase curve information.  The first column again gives the name (or designation) of the icy body in the same order as appearing in Tables 1 and 2.  The second column gives the slope of the phase curve at zero phase angle as averaged over all bands, $S_0$ in units of magnitudes per degree of phase.  This averaging is to improve the statistical error bars.  Values are quoted only for those bodies with phase curves with useful data below 0.5$\degr$ phase.  In principle, as phase curves are never measured down to zero phase, an unobserved sharp upturn in the phase curve could make for substantial errors in $S_0$.  However, as discussed later, such hidden 'spikes' can only arise from the coherent backscattering mechanism, and are normally present on bodies with a large unobserved range of low phases, and for which the observed phase curve is nearly flat.  Thus, in practice, this concern is only significant for Pluto and 120178=2003 OP32.  The third column gives the slope of the phase curve at a phase of $1\degr$, $S_1$.  The reason to provide slopes at two separate phases is that some phase curves show nonlinear shapes (in particular, a sharp steepening towards zero phase which can be called an 'opposition spike').  Our choices of $0\degr$ and $1\degr$ are because these phases are essentially observable for almost all of our 52 bodies.  Almost all of the non-moon icy bodies have phase curves that are consistent with a linear shape, so the average slope applies to the entire region of phase angle from roughly $0\degr$ to $2\degr$, and we set $S_0=S_1$ if no significant slope is detected.  As discussed in RST2007, the phase curves for Thereus and Bienor might be better fit with a non-linear curve, but the linear slope nevertheless provides a reasonable description of the phase curve.  Most of the moons display a decidedly non-linear phase curve.  The fourth column gives the magnitude difference in the phase curve between phases of $2\degr$ and $0\degr$, $\Delta m_{0-2}$ in units of magnitudes.  In some cases, small extrapolations (to either $0\degr$ or $2\degr$ phase) have been made, and there is potential for some small error to arise from this.  In some cases (e.g. where the phase curve can only be measured from the ground out to a phase angle of $1.4\degr$), the extrapolation to $2\degr$ is fairly far, in which case we have presented the $\Delta m_{0-2}$ value as a range bounded by that from a simple linear extrapolation of the phase curve and from an assumption that the phase curve goes flat for unobserved phases.  In the case of Pluto and Charon, the phase curve is closely linear over the phase range of $0.57\degr-1.94\degr$, so the quoted $\Delta m_{0-2}$ value is based on the linear fit even though a very narrow opposition surge could conceivably raise this value somewhat.  If the possible phase range is too small to provide coverage over a substantial portion of the $0\degr-2\degr$ range, then we do not give any value for $\Delta m_{0-2}$.  The fifth column gives the numerical difference between the phase curve slopes as measured in the B-band and the I-band (in magnitudes).  For some bodies, we do not have B-band and I-band phase curves, but instead have other bands, and if on this basis we see that the slope does not vary significantly with color and then these are represented as "$\sim0.00$".  The last column gives the source for the measured surge properties.
	
\section{The Cause of the Opposition Surges}

	The solar phase angle ($\phi$) is the Sun-KBO-Earth angle, with $\phi=0 \degr$ corresponding to opposition.  The opposition surge is a non-linear brightening as the phase angle gets close to zero.  The phase curve is a plot of the KBOÕs magnitude versus the phase, with an upward slope getting steeper as $\phi$ approaches $0 \degr$.  The surge has two distinct physical mechanisms.  First, shadow-hiding (SH) results from the shadows being hidden behind the objects casting the shadows, so that at opposition no shadows are seen and the target appears brighter than when the phase is larger and shadows darken the target.  Second, coherent backscattering (CB) is an interference effect resulting from light rays following the same path - only backwards - that constructively interfere with each other resulting in a doubling of that light's intensity.
	
	SH is a mechanism that has been long known and is easily observed.  The properties of the SH surge depend on the texture of the surface.  The surfaces might be jagged in shape or there might be opaque particles embedded in clear ices, with both casting shadows on the lower surface.  Viewed from the side (at ÔhighÕ phase angles), these shadows are visible, so that the overall surface will consist of bright sunlit areas and dark shadows.  Viewed from zero phase angle, the viewer will not see the shadows, as they will all be hidden behind the objects casting the shadows.  Detailed physical models for SH are presented in Hapke (1986; 1993; 2002) and Shkuratov \& Helfenstein (2001).
	
	As a surprisingly late realization, an entirely new effect was discovered to provide a second mechanism for an opposition surge.  What is going on at zero phase is that double-scattered light can follow some path and its reverse, such that for closely positioned scatterers the two paths will constructively interfere with each other resulting in a doubling of the intensity.  As the phase angle moves away from zero, the constructive interference decreases.  Detailed physical models for SH are presented in Hapke (1993; 2002) and Shkuratov \& Helfenstein (2001).
	
	From the time that it was realized that two different surge mechanisms were possible, a primary question has been to distinguish which mechanisms are operating.  Likely, for realistic surfaces, both mechanisms are operating, so the question then becomes as to which mechanism is dominating, with the answer being dependent on the phase region being discussed.  The correct identification of the amplitude of each component is required for any physical analysis of the observed surges.  An analysis of the phase curves can yield the physical properties of the surface (like porosity, the mean free path, and the single particle albedo), but only if we know whether to apply models for SH or CB.  Currently, theory does not know how to predict the amplitudes for either mechanism, so this knowledge must come empirically from observations.
	
\subsection{Criteria for Distinguishing Surge Mechanisms}

	How can we distinguish observationally the dominant surge mechanism?  Two of the early criteria have had to be revised later.  First, Hapke (1993) originally gave the prediction that the width of the CB surge component should scale linearly with wavelength, and this indeed provided our first basis for having multiple bands widely spaced (B, V, and I) for our SMARTS observations.  It was later found that CB widths can in fact have weak dependence on wavelength (Nelson et al. 2000).  Second, CB was expected to be minimal for bodies with low albedo as there should be little multiple scattering as required for the CB effect.  It was later found however that low-albedo bodies can have strong CB (Hapke, Nelson, \& Smythe 1993; 1998; Helfenfenstein, Veverka, and Hillier 1997).  Theoretical consideration of scattering within a single particle shows that CB can be strong for light that is effectively singly scattered (Hapke 2002).
	
	The old presumption that opposition surges are caused by SH has been successfully challenged in a few cases.  The first and best case has been for our Moon, for which the observed surge turns out to be roughly equal amounts of SH and CB (Hapke, Nelson, \& Smythe 1993; 1998; Buratti, Hillier, \& Wang 1996; Helfenstein, Veverka, \& Hillier 1997; Shkuratov et al. 1999).  In this case, the CB contribution was demonstrated from laboratory measures of lunar soil samples, from a small wavelength dependence, and from the polarization as a function of phase.  The exact shape of the phase curve has been used to measure the SH and CB contributions for Saturnian moons and rings that passed through true opposition in 2005, with the result that CB dominates at low phase angles (for $\phi \lesssim 1 \degr$) and SH dominates at relatively high phase angles (French et al. 2007; Verbiscer et al. 2006).  Our group has also used the exact shape of the phase curve to distinguish CB from SH for Nereid, with the result that the surge for $\phi < 2 \degr$ is dominated by CB (Schaefer et al. 2007).
	
	We have constructed four separate criteria for distinguishing between SH and CB.  Here, we focus on readily measurable quantities for phase ranges accessible from Earth (typically $\phi < 2 \degr$).  Additional criteria are possible, based for example on polarization (Bagnulo et al. 2006), but such polarization measures are available for few KBOs or Centaurs.
	
\subsubsection{Color-Independence of Surge}

	SH is nearly color independent because the positions of the shadows does not depend on the wavelength of the light.  CB can make the phase curve change with wavelength, even though this dependence might be weak.  This sets up a simple criterion for distinguishing between SH and CB; if the observed phase curve varies significantly with color, then the surge must have a significant CB component.  Generally, if the effect is large enough to be significant, then the CB component must dominate over the SH component.  This criterion does not work in reverse, as an apparently color-independent surge might either be dominated by SH or by a CB component that happens not to vary much with wavelength.
		
		In practice, this criterion can be expressed by comparing the low-phase-angle slopes of the phase curve in B-band and I-band.  The slopes of the phase curve are $S_B$ and $S_I$ in the B- and I-band respectively at some phase angle, and the uncertainty in $S_B-S_I$ is $\sigma _{B-I}$.  The fifth column of Table 3 presents $S_B-S_I \pm \sigma_{B-I}$.  If the blue slope differs from the red slope by more than about two-sigma, then we can be confident that the surge is being dominated by CB.  Thus, our first criterion is that CB dominates if $\vert S_B-S_I \vert /\sigma_{B-I} > 2$.
		
		The existence of albedo changes with wavelength will complicate this analysis somewhat.  For example, for the SH mechanism, the albedo in the I-band is higher than in the B-band, so the shadows will be more filled in for red light leading to a lower amplitude and hence a smaller slope in the red so that $S_B-S_I$ will be slightly positive.
		
		Table 4 presents the results of our four criteria for all 52 icy bodies.  The first column gives the name, in the same order as in the first three tables.  The second column gives the conclusions based on our first criterion.  If $\vert S_B-S_I \vert /\sigma_{B-I} > 2$, then the dominant cause of the surge cannot be SH, so our conclusion is that the dominant cause is coherent backscattering as represented by a 'CB'.  In three cases (29981=1999 TD10, 73480=2002 PN34, and Haumea), this criterion is formally satisfied, yet the slope difference is so small that we acknowledge that even small unrecognized errors will change the conclusion, so we give our result as "?".  If $\vert S_B-S_I \vert /\sigma_{B-I} < 2$, then the surge could be caused by either SH or CB.  We leave as blank those cases for which we do not have information on the slope differences.  In all, we identify seven icy bodies for which we can confidently conclude that the dominant cause of the surge is CB.
			
\subsubsection{Slope of Phase Curve}

	Both SH and CB mechanisms have characteristic angular widths in the phase curve, designated as $h_S$ and $h_C$ respectively.  For SH, the width is determined by the typical aspect ratio of the shadow, which is the characteristic size of the shadow casters divided by the length of the shadow being cast.  For ordinary icy surfaces, the aspect ratio cannot be large, and this puts an effective lower limit on possible values for $h_S$, with $h_S \gtrsim 6\degr$.  (This limit is set by the aspect ratio of any typical jumbled geometry, where the aspect ratio is the typical size of the shadow casters divided by the length of the shadow.  In principle, this limit could be violated for special geometries 'pointing at' Earth or in the case of low density scatterers suspended in clear ice.)  This limit is also returned from theoretical models of packed particles (Hapke 1993) and for icy bodies with accurate phase curves over wide phase ranges (e.g., Verbiscer, French, \& McGhee 2005).  For CB, the width is determined by the relative spacing of the scattering sites compared to the wavelength of the light.  Both experimentally and observationally, the typical values of width have $h_C \lesssim 3\degr$.  This distinction in possible surge widths provides a means to distinguish the dominant mechanism even with moderate or poor phase curves.
	
	The slope of the phase curve at low phase angles depends on the mechanism, the surge width, and the amplitude of the surge.  Detailed functional forms are given by Hapke (2002).  The maximal slopes for both mechanisms will occur at zero phase angle with maximal amplitude.  The amplitudes for the two mechanisms, designated $B_{S0}$ and $B_{C0}$, both have maximal values of 1.0, which corresponds to a doubling of the light's intensity at zero phase angle.  For SH, the low phase angle slope is $S_0 \approx 0.2(B_{S0}/h_S)$ mag deg$^{-1}$.  Given that $B_{S0} \leq 1$ and $h_S \gtrsim 6\degr$, we have that the steepest slope in a phase curve that is possible for the SH mechanism is $S_0 \approx 0.033$ mag deg$^{-1}$.  Thus, if we see a slope in a phase curve that is significantly steeper than 0.033 mag deg$^{-1}$, then we know that the CB mechanism is dominant.  This criterion does not work in reverse, as a shallow phase curve slope could arise from any combination of several cases; the SH mechanism dominates, the CB mechanism dominates but with a low $B_{C0}$ value, the CB mechanism dominates with a small $h_C$ producing a narrow opposition spike that might be apparent only at unobserved low phase angles, or that neither mechanism is operating.
	
	In practice, we have to allow for the normal measurement uncertainties in the average slope at low phase angles.  The second column of Table 3 presents $S_0 \pm \sigma_{S_0}$.  If the average slope is greater than 0.04 mag deg$^{-1}$ by more than two-sigma, then we can be confident that the CB mechanism dominates.  So our second criterion is that CB dominates if $S_0-2\sigma_{S_0} > 0.04$ mag deg$^{-1}$.
	
	The third column in Table 4 presents our results for this second criterion.  The icy bodies which satisfy our second criterion are designated with 'CB'.  The bodies which do not satisfy this criterion could have either SH or CB dominating.  In all, we identify 41 icy bodies for which we can confidently conclude that the dominant surge mechanism is CB.
	
\subsubsection{Shape of Phase Curve}

	Hapke (2002) presents a detailed physical model for the phase curve with both the SH and CB effects included.  For the case where we are dealing with small phase angles, $\phi$, the observed magnitude will be 
\begin{equation}
m = m_0 - 2.5 \times \log _{10}[p(1+B_{S0}B_S)+M] - 2.5 \times \log _{10}[1+B_{C0}B_C], 
\end{equation} 		
\begin{equation}
B_S = (1+[\tan (\phi/2)]/h_S)^{-1},
\end{equation} 		
\begin{equation}
B_C = 0.5(1+[(1-e^{-Z})/Z])/(1+Z)^2,
\end{equation} 		
\begin{equation}
Z=[\tan (\phi/2)] / h_C.
\end{equation} 		
Here, $m_0$ is a magnitude that has absorbed various constants, $p$ is the single particle scattering function evaluated for $\phi \sim 0 \degr$, and $M$ is a constant that represents the multiple scattering contribution.  The contribution to the surge from SH is given by $B_S$ with an amplitude of $B_{S0}$ and a width $h_S$.  The contribution to the surge from CB is given by $B_C$ with an amplitude of $B_{C0}$ and a width of $h_C$.  The two amplitudes must be between zero and one, so that both the SH and CB effects individually can at most double the brightness of the body (i.e., an increase by 0.75 mag).  To get amplitudes of near unity requires contrived conditions, so in practice the amplitudes are never much larger than a half magnitude or so.  The angular widths have to be expressed in radians for equations 2 and 4, although in the text we will express the widths in the more convenient unit of degrees.  The parameter $Z$ is separated out of equation 3 for simplicity of representation.
	
	Our point for displaying equations 1-4 is that the SH and CB phase curves have different {\it shapes}, and that this can be used to distinguish between the two mechanisms.  That is, for sufficiently accurate magnitudes spread over a wide-enough range of phases, the observed phase curve can be fit so as to find $B_{S0}$ and $B_{C0}$ which will then tell us the relative contributions.
	
	With characteristic surge widths of $h_S \gtrsim 6 \degr$ and $h_C \lesssim 3\degr$, the high phase angles will be where SH is apparent and the low phase angles will be where CB is apparent.  For Centaurs, with their orbits inside Neptune, we can observe out to phases of $4\degr$ to $8\degr$, and both surge components might be apparent.  Unfortunately, for KBOs, their maximum observable phase is $\sim2\degr$ (to as low as $0.46\degr$ for Sedna) and this makes it difficult to untangle the relative contributions of the two components.
	
	For top quality phase curves, we can fit equations 1-4 and perhaps quantify the relative contributions of both components or perhaps at least distinguish whether one component is required or rejected.  For this test, the fitted error bars on the amplitudes are critical for knowing the confidence level of any conclusion.  So our third criterion is to fit the shape of the well-measured phase curves and look at the amplitudes.
	
	Nereid provides a good example of the use of this criterion.  We use our data from Schaefer et al. (2007), and select out only the years 1998, 1999, 2000, and 2006 for which the intrinsic variability was minimal (see Figure 1).  This presents a well-measured phase curve for which we can distinguish between the shape of the SH and CB surges.  We fitted the data to various models (see Table 5), including ones with SH only (either with $h_S$ unconstrained or fixed at $6\degr$), CB only, or SH-plus-CB.  We first see  that the SH-only fit with $h_S$ pegged to the smallest plausible value is much too flat and completely unable to explain the surge.  If we allow the $h_S$ value to be unconstrained, then we get the best fit for unlikely values with $h_S=0.20\degr$.  This specific strong rejection of the SH model for Nereid (that $h_S$ must be much smaller than is possible) is really just a restatement of our second criterion based on the slope at small phase angles.  Our third criterion regarding the shape of the phase curve comes from a comparison between the SH-only and the CB-only fits.  The SH-only fit has a chi-square of 422.0 versus 339.1 for the CB-only fit (with 194 degrees of freedom in the fit).  This huge difference in chi-square is the proof that the CB mechanism dominates the observed surge.  Nevertheless, just because the CB mechanism dominates does not mean that there is no shadow hiding.  In particular, the chi-square is improved by having both SH and CB, although the improvement is small compared to the difference between the SH-only and CB-only fits.  With the model differences between the CB-only and SH+CB fits being smaller than 0.006 mag over the entire range of observed phase angles, we realize that the SH component is poorly constrained and can at most exert a slight tilt in the phase curve.  The reduced chi-square for the 196 data points is 1.71, and this is likely due to residual variations of Nereid even in its quietest years.  In all, the shape of Nereid's phase curve demonstrates that CB is the dominant mechanism (even allowing for unphysically small widths for SH).
	
	A well-measured phase curve is required to be able to make the distinction between the shapes of the two mechanisms.  Verbiscer et al. (2005) have a top quality multi-color phase curve for Enceladus down to phase angle $0.26\degr$ with {\it Hubble Space Telescope} observations.  Their Figure 4 exemplifies the difference between the shapes, with the largest differences being at the lowest phase angles.  The CB contribution is quite steep at low phases ($\phi \lesssim h_C$) and is nearly flat at high phases.  With further ground-based data extending to zero phase angles, the surge of Enceladus is seen to rise sharply towards true opposition (Verbiscer 2007, private communication), and this points to the dominance of CB at least at low phase angles.  Another case with good data and a detailed analysis is for Titania, again with the conclusion that CB dominates at small phase angles (Shkuratov \& Helfenstein 2001).  Indeed, for the six Uranian satellites, Karkoschka (2001) demonstrates that CB is required to get the observed brightening for phase angles under one degree.  Another good case is Europa as viewed with the {\it Galileo} spacecraft, where the surge properties are resolved across small surface features and the CB surge dominates at low phase angles with a surprisingly narrow ($\sim 0.2\degr$) width (Helfenstein et al. 1998).
	
	The fourth column in Table 4 presents the results for our third criterion.  These results are all taken from detailed analyses of the phases curves from the references cited in Table 3.  Unfortunately, only 9 of our 52 icy bodies have adequate data and analyses.  All 9 of these icy bodies have phase curve shapes that demonstrate that CB dominates at low phase angles.

\subsubsection{Albedo}

	SH only works for single scattered light, as multiple scattering will fill-in the shadows.  Low-albedo surfaces will have single scattering dominating over multiple scattering, because the intensity of the scattered light is reduced by a factor equal to the albedo for each scattering.  High-albedo surfaces will have much multiple scattering which will fill-in the shadows and eliminate the SH mechanism.   The CB mechanism requires multiple scattering such as is present on high-albedo surfaces.  However, the CB effect is also now known occur in moderate- and low-albedo surfaces.  If the surface has a high albedo, then the surge must be dominated by CB.  This criterion does not work in reverse, as a low-albedo surface could have either SH or CB dominating.
	
	To measure the geometric albedo ($p_V$), we must have the brightness and radius of the body.  The radius is hard to get for many bodies.  For moons of the planets, we have resolved images which will give the radii.  For Pluto and Charon, we have the mutual transit/eclipse events of the 1980s as well as later stellar occultations to provide sizes.  For other bodies, we can only try to measure the brightness in the thermal infrared so as to get a simultaneous solution to the radius and albedo.  Fortunately, with $24 \mu$ and $70 \mu$ thermal infrared brightness measures with the {\it Spitzer Space Telescope}, Stansberry et al. (2007) has derived radii and albedos for many of our KBOs (see Table 2).
	
	In practice, we have to estimate some dividing line above which the SH cannot dominate.  This dividing line will have to be somewhat uncertain as it will depend on unknown circumstances and the degree of domination required.  Shkuratov \& Helfenstein (2001) demonstrate that the CB is already dominating by a factor of seven for single particle albedos of 50\%.  So an albedo of 40\% will certainly have CB domination by a significant factor, even though our dividing line could be pushed lower.  This threshold is not critical, as it makes no difference to any of the conclusions in Table 4 as long as the threshold is set anywhere from 10\% to 60\%.  So our fourth criterion is that CB dominates if $p_V > 40\%$.  Nevertheless, a high albedo does not guarantee that the CB surge will have a substantial amplitude, as for example fresh snow has an essentially flat phase curve.
	
	The fifth column in Table 4 presents our results for our fourth criterion.  If $p_V > 40\%$, then the shadows are filled and CB must dominate.  If $p_V < 40\%$, then either SH or CB could dominate.  For bodies with no available albedo or only ambiguous limits, we have left the conclusion blank.  In all, we identify 12 icy bodies for which the high albedo gives us confidence that CB dominates.
	
\subsection{Results}

	Table 4 gives the results from all four of our criteria for distinguishing whether SH or CB dominates the observed phase curves.  All four criteria happen to work such that we can confidently identify CB based on some condition, whereas if the condition is not satisfied then we cannot decide which mechanism dominates.  So if any one of our criteria is passed then we conclude with confidence that CB dominates, whereas if none of our criteria are passed (perhaps due in part to inadequate observational material) then we are left with no confident conclusion as to the mechanism.  With our criteria unable to point exclusively to the SH mechanism, we cannot have any contradiction between criteria.  We find almost all of our 52 icy bodies pass one or more of our criteria and we can thus be confident that their surges at low phase angles are dominated by CB.  We note that Tethys and Dione both have small surges, but their high albedo requires that the small surges are dominated by CB.  These conclusions are listed in the last column of Table 4.  We find only four bodies (73480=2002 PN34, Asbolus, 24835=1995 SM55, and 55636=2002 TX300) that have no indication as to the surge cause.  We also find two bodies (95626=2002 GZ32 and 120178=2003 OP32) which have essentially no apparent surge. 
	
%	We have found that nearly all the small icy bodies in the outer Solar System with surges have the CB mechanism dominating at low phase angles.  This is a fairly sweeping result.  Nevertheless, it is worthwhile to point out small limitations in the generality of our conclusion.  First, 6 out of 52 icy bodies either do not have a measurable surge or do not have data to prove that the surge must be dominated by CB.  A reasonable counter to this limitation is that the 6 bodies might well have CB-dominated surges even if they are small or not adequately measured.  Second, the surge at relatively large phase angles (say, $>3\degr$) can still be dominated by the SH mechanism.  A reasonable counter to this limitation is that the phase curve slope at high angles must necessarily be shallow and should not be called surges, certainly not in comparison with the sharp CB spikes at low phases.  Third, SH enhancements may still be present at low phase angles despite the relatively steep slopes being caused by CB.  A reasonable counter to this limitation is to agree that SH might be present, but to recall that the conclusion is regarding which mechanism dominates and the SH does not dominate.  Thus, despite these small limitations, we are still left with the conclusion that CB dominates on nearly all the icy bodies in the outer Solar System.
	
\section{Hapke Surge Parameters}

	With Hapke's model for the opposition surge, both mechanisms have a width and an amplitude.  In principle, these can be derived from a formal fit to the phase curve data.  The equations relevant for fits restricted to a narrow range of near-zero phase angles are simply the equations 1-4 above.  We have made formal chi-square fits for ten of our phase curves, using data from RST2007, Schaefer et al. (2007), and Schaefer \& Rabinowitz (2002).  All of the fits (other than for Nereid) were performed on binned phase curves, so the number of points is greatly smaller than the number of individual observations.  We have set limits that the amplitudes must be $\leq1$, $h_S>6\degr$, and $h_C \lesssim 3\degr$.  In a number of these fits, some of the parameters are pegged up against their limits, as indicated by the superscript 'pegged' in the Table.  The results of these fits are presented in Table 6.  For all ten objects, we see that the CB effect has $0.6\degr < h_C \lesssim 3\degr$ and $0.4 \lesssim B_{C0} \leq 1$.
	
	In these fits, the differences in chi-square between the CB-only and the SH-plus-CB fits are insignificantly small, which is to say that the maximum allowable SH effects cannot substantially change the shape of the phase curve.  Another way of saying this is that the CB mechanism dominates and the SH mechanism might or might not be present.  The only exception to this is Nereid, where the addition of a SH component makes an improvement in the fit of moderate significance (see Table 5).
	
	Unfortunately, many of the phase curves are not of sufficient accuracy to appear to warrant any formal fits.  Here, we will put forward an approximate method to get the surge parameters in a simple way directly from the phase curve.  Our method involves the extraction of two observable properties that are easily taken from the phase curve, $S_0$ and $\Delta m_{0-2}$.  One is the slope at low phase angles and the other is the difference between two phases.  We have already tabulated these two parameters for our 52 objects in Table 3.  These have been plotted in Figure 2a.
	
	The theoretical curves from Hapke's model can be calculated from equations 1-4.  For each set of surge parameters, we can calculate the phase curve and derive the values for $S_0$ (taken as the slope over the phase range from $0.0\degr$ to $0.1\degr$) and $\Delta m_{0-2}$.  In Figure 2b, we display the theoretical curves for both SH and CB as their widths and amplitudes change.  
	
	For SH, with $h_S>6\degr$ and $B_{S0} \leq 1$, the allowed region is forced into a very small region in the lower left corner of the plot.  That is, we always have $S_0<0.029$ mag deg$^{-1}$ and $\Delta m_{0-2}<0.052$.  We immediately see that almost all the icy bodies certainly have much too large a surge to be explainable by SH.  There are five objects that are consistent with a SH dominated surge.  Of these, three have their albedo $>50\%$ and must have their small surges dominated by CB.
	
	For CB, with $h_C<3\degr$ and $B_{C0} \leq 1$, the surges can be large in the $0\degr-2\degr$ phase range.  The $\Delta m_{0-2}$ value can never be larger than 0.75 mag (i.e., a factor of 2) and it approaches this value only with $B_{C0} = 1$ and narrow surges.  Figure 2 displays several curves for $h_C$ varying from $0.1\degr-3.0\degr$, with $B_{C0}$ values of 1.0, 0.5, 0.25, and 0.1.  Figure 2b also displays several curves for $B_{C0}$ varying from $0.1-1.0$, with $h_C$ values of $0.2\degr$, $1.0\degr$, and $3.0\degr$.  On this diagram, there is a forbidden region below and to the right of the $B_{C0}=1$ curve.
	
	In principle and in practice, an icy body can have both SH and CB.  But the effects of SH are always small and little effect the CB-only curves.  For example, even with extreme SH effects included ($h_S=6\degr$ and $B_{S0}=1$), the $\Delta m_{0-2}$ differs by only 0.05 mag while the $S_0$ value differs by only 0.03 mag.  Such effects are generally smaller than the photometric uncertainties and are hence negligible.
	
	From Figures 2a and 2b, we can read off the surge properties for the icy bodies.  The position of Nereid shows that $h_C \approx 0.9\degr$ and $B_{C0} \approx 0.6$, which is in good agreement with the results of a formal fit for CB only (see Table 1).  The high-albedo Uranian moons Ariel, Titania, and Oberon all have $h_c \sim 0.3\degr$ and $B_{C0} \sim 1$.  The high-albedo Neptunian moon Triton has $h_c \sim 0.05\degr$ and $B_{C0} \sim 0.1$.  The rest of the objects all are close to the line ($S_0=0.5\Delta m_{0-2}$) which corresponds to a straight-line phase curve.  As such, the only way for this to occur is for the CB width to be near its maximum value of $3\degr$.
	
	A concern that we initially had was that several objects (notably Varuna, 47171=1999 TC36, and 47932=2000GN171) are inside the forbidden region (to the right and below the $B_{C0}=1$ curve in Figure 2a).  An additional case would be the I-band phase curve for Quaoar.  The problem is that the phase curves are apparently too straight for their large amplitude.  The formal error bars in Figure 2b easily allow for a resolution of our concern as the observed phase curves have uncertainties large enough to allow for an acceptable amount of curvature.  We have tested whether any of these phase curves are inconsistent with the Hapke model by means of a formal fit.  These fits are presented in the last four lines of Table 6.  Not surprisingly, the $B_{C0}$ values have pegged at their maximum value of 1.  However, the chi-square values are reasonable for a good fit, which indicates that the Hapke model is acceptable even in these worst cases.  That is, despite being in the 'forbidden region' of Figure 2a, the uncertainties in the phase curves are such that the required curvature is still allowed and so our original concern is not realized.
	
	Another concern that we have is that some bodies might have a very narrow opposition surge (an 'opposition spike') at low phase angles where there might not be any observations.  (For our data, the cause for the lack of low-phase data is only that the body's inclination makes the low-phase-angles to be unobservable from Earth.)  Such a narrow spike can only occur from the CB mechanism.  Bodies with a phase curve slope of $>0.04$ mag deg$^{-1}$ already have a CB surge with a width that is too wide to allow for the existence of a spike.  (In principle, the surface could have two regions, one with a wide CB surge and the other with a narrow CB surge, such that a spike could ride on top of a normal wide surge, but this possibility seems contrived and has never been observed on any body.)  Thus, the only way for an unobserved spike to be hidden is if the observed phase curve is nearly flat and a substantial range of low phase angles remain unobserved.  This case would be as for the narrow spike in the phase curve of Triton (Buratti, Bauer, \& Hicks 2007) had there been no observations inside a phase of 0.15$\degr$.  A prominent possible case for a missing spike is Pluto, for which the flat phase curve in Buie, Tholen, \& Wasserman (1997) only extends in to a phase of 0.57$\degr$.  (The Centaur 95626=2002 GZ32, the Large body Makemake, and the two Collisional bodies 145453=2005 RR43 and 120178=2003 OP32 might possibly have spikes also.)  Pluto could well have a narrow CB spike with full amplitude ($B_{C0}=1$) and a narrow width ($h_C \lesssim 0.1 \degr$).
		
\section{Surge Correlations}

	With our new database of surge measurements, we can start looking for correlations of surges with specific surface properties.  In this section we report three new correlations, all of which are highly significant.
	
\subsection{Surge-Color Correlation}

	Figure 3 shows a plot of the surge slope ($S_1$) versus the color ($B-I$).  The planetary moons have not been included because they have many surface reprocessing mechanisms and unique histories so that their surface properties could well be widely varied.  (Indeed, the moons are widely scattered on the slope versus $B-I$ plot.)  The non-moon bodies show a general correlation, with redder bodies displaying steeper surge slopes.  This correlation is rather wide, yet nevertheless there is a striking lack of any bodies in the lower-right and upper-left portions of the diagram.  That is, for some reason, nature is not making relatively-blue bodies with steep surge slopes and is not making reddish objects with shallow-slope surges.  A correlation line is also displayed in Figure 3, with this being the bisector slope (Isobe et al. 1990) with an equation of $B-I=1.35+4.8S_1$.  The usual correlation coefficient (r) has a value of 0.51 for 37 surge-color pairs.  The probability that uncorrelated pairs would produce a correlation coefficient with $\mid r \mid \ge 0.51$ is 0.0012.  That is, the existence of the correlation is highly significant, even if the scatter is fairly large.
	
	The individual bodies within each  group do not appear to follow the correlation.  (The Small/Red bodies might follow the correlation line, but we judge this to not be significant for the data in hand.)  So the correlation is created by the placement of the various groups, notably with the Small/Red bodies in the upper-right of the plot and the Collisional and Large bodies to the lower-left of the plot. 
	
	Belskaya and Shevchenko (2000) have found a similar surge-color correlation for asteroids.
	
	We can speculate on the cause the the surge-color correlation.  We think that the oldest surfaces have the reddest colors and the steepest slopes while the youngest surfaces have the bluest colors and the shallowest surges.  Apparently, the surge slopes and colors are both age indicators for the surface.  For the colors, we already know that radiation exposure will slowly turn icy surfaces redder in color, so the older the surface the redder the color.  Indeed, the oldest surfaces (the Small/Red bodies) are in the upper-right of the diagram.  The youngest surfaces (including most of the Intermediate and Large bodies with their seasonal frosts and cryovolcanism) are in the lower-left of the plot.  The Collisional bodies are also in the lower left, perhaps as appropriate for their age being relatively young.  In all, we think that the correlation in Figure 3 is caused by the increasing age of the surface.
	
\subsection{Surge-Albedo Correlation}

	We note a second strong correlation in our surge data base, and that is between $\Delta m_{0-2}$ and $p_V$ for the bodies with icy surfaces.  (The Small/Gray bodies do not follow this correlation and they are excluded as having largely ice-free surfaces.)  Figure 4 shows a plot between these two quantities.  In several cases, the $\Delta m_{0-2}$ were evaluated by assuming a linear fit over the entire phase range.  We see that all the high albedo bodies (with $p_V>40\%$) have low surges (with $\Delta m_{0-2} \leq 0.25$ mag), or alternatively that all high surge bodies (with $\Delta m_{0-2} > 0.25$ mag) have low albedos (with $p_V \leq 40\%$).  Another way to look at this correlation is to realize that there is a triangular shaped empty region in the plot (connecting the two points on the axes with $\Delta m_{0-2}=0.22$ mag and $p_V=70\%$) such that there is an absence of low-surge low-albedo bodies.  This correlation is highly significant, even though we can see substantial scatter in the relation.
	
	Belskaya and Shevchenko (2000) have found a similar effect for the asteroids and meteorites.  Laboratory measures of chalk/soot mixtures show a relatively complex behavior where the phase curve steepens over a smaller-and-smaller phase range near zero as the albedo increases, but that the slope at large phases flattens somewhat as the albedo increases (Shkuratov et al. 2002).  Laboratory measurements (Nelson et al. 2004) also show that the surge albedo correlation is not monotonic, with the CB effect rising to a maximum at albedos of up to 40\% (as the degree of multiple scattering increases) but then declines as the albedo increases (due to the increasing number of scatterings having the effect of randomizing the polarization and weakening the CB effect). 
	
	We speculate that our observed surge-albedo correlation in ices arises because both high albedo and flat phase curves are produced by recently reprocessed icy surfaces.  Shkuratov et al. (2002) finds a very flat phase curve for freshly fallen snow.  In Figure 4, many of the bodies occupying the upper-left corner are ones with ongoing snow or frost formation due to tenuous atmospheres, cryovolcanism, or water extrusion (Pluto, Eris, Triton, Makemake, Haumea, Europa, and Enceladus).  In contrast, the many bodies in the lower-right part of Figure 4 are the Small/Red bodies (plus Quaoar) with no recent frost or snow deposition.  Even with the correlation between the age of the surface and the position in Figure 4, we acknowledge that the scatter implies that the real situation is more complicated than the simple part that we have just put forward.
	
\subsection{Surge-Inclination Correlation}

	Tegler, Romanishin, \& Consolmagno (2003) and Peixinho et al. (2004) have found that KBO color distributions depend on whether the object has high or low inclination.  Knoll et al. (2007) discovered that the fraction of binary KBOs is 29\% for objects with inclinations smaller than 5.5$\degr$, while the binary fraction is 9.5\% for objects with higher inclinations.  Such correlations are presumably related to the frequency and velocity of collisions amongst the KBOs.  So it is reasonable to seek surge-inclination correlations.  To this end, Figure 5 has a plot for our 36 non-moon bodies.
	
	In Figure 5, we see a large empty region in the lower-left side of the plot.  This empty region is for $i<15\degr$ and $S_0< 0.12$ mag deg$^{-1}$.  For $i\ge15\degr$, we see that 16 out of 24 have $S_0 < 0.12$ mag deg$^{-1}$ (67\%).  For $i<15\degr$, zero out of 13 have $S_0 < 0.12$ mag deg$^{-1}$ (0\%).  For ordinary binomial probabilities, this is extremely improbable by chance alone.  As such, we conclude that the low inclination KBOs never have shallow surges.  Alternatively, we could say that all bodies with low surge slopes have high inclination.
	
	Now we have a suite of three properties that are tied in to the orbital inclination; surge slope, color, and binary frequency.  The low inclination objects all have steep surges, red colors, and frequent binary companions.  The high inclination objects have a full range of surge slopes, a full range of colors, and a low frequency of binary companions.
	
	We speculate that the origin of this surge-inclination relation is that the low inclination bodies have older surfaces (and hence have redder colors and steeper surges), while the higher inclination bodies have suffered orbital perturbations from their original low-inclination orbits.  The higher inclination bodies would have lost their satellites during whatever perturbation history lead to their orbits changing to high inclination.  Some of these low-inclination bodies scattered to higher inclination might pass through a period during which they get closer to the Sun or have collisions, either of which could change their surfaces to low surge slopes and gray colors.  This speculation has difficulty explaining why all the low inclination Centaurs have relatively high surge slopes.  An alternative speculation is that the high- and low-inclination bodies have different source populations (Tegler, Romanishin, \& Consolmagno 2003).
	
\section{Nereid}

	Nereid is one of the most mysterious objects in our Solar System.  For example it experiences secular changes in its variability from high-amplitude to low-amplitude (e.g., Schaefer et al. 2007).  Another mysterious aspect is its origin.  Plausible ideas include that it is a captured Kuiper Belt Object or an almost-ejected inner moon of Neptune.  The surface properties can be used to help distinguish between these origins.
	
	Our phase curve for Nereid shows a steep surge at low phase angles and a very large surge amplitude between $2\degr$ and $0\degr$.  These properties are greatly different from all Kuiper Belt Objects (including Centaurs).  In addition, the albedo of Nereid (40\%) is greatly higher than all candidate classes of captured objects.  Also, the color of Nereid is outside the observed range of all groups other than the Haumea collisional family.  As such, we can conclude that Nereid is {\it not} a captured KBO.  The only loophole to this rejection is if Nereid has undergone substantial surface reprocessing (that no other KBO undergoes) despite the severe isolation provided by its one year orbital period around Neptune.
	
	As for the other plausible idea, that Nereid is a nearly-ejected inner moon of Neptune, we see good agreement.  That is, the surge properties, albedo, and color fit in well with other small inner icy moons.  For example, Nereid has surface properties that are not substantially different than those of the Uranian moons.  In all, based on the observed surface properties, Nereid is apparently a former inner moon of Neptune.
	
\section{Conclusions}

	Our group has provided for the first time a large number of well-measured phase curves for non-moon icy bodies in the outer Solar System (RST2007; Schaefer et al. 2008; Schaefer \& Rabinowitz 2002; Schaefer \& Tourtellotte 2001; and this paper).  To this large data set, we have added the results from the literature for five phase curves on Centaurs and KBOs plus phase curves for 16 icy moons.  With our compilation of surge, surface, and orbital properties for 52 icy bodies, we have addressed two broad questions.  The first is regarding the cause of the surge, which is to say that we seek to identify the dominant surge mechanism (between SH and CB).  The second is that we seek correlations between the surge properties and the other surface and orbital properties.  Here are our conclusions:
	
	(1) We give four specific criteria based on simple observable properties that can prove that the dominant surge mechanism is CB.  The CB mechanism must dominate if the surge slope is color dependent ($\vert S_B-S_I \vert /\sigma_{B-I} > 2$), if the surge slope is significantly steeper than allowed by SH ($S-2\sigma_S > 0.04$ mag deg$^{-1}$), if the shape of the phase curve matches the CB model (Equations 1-4), or if the albedo is sufficiently high such that the shadows get filled in ($p_V > 40\%$).
	
	(2) We find that the CB mechanism dominates at low phase angles for virtually all of our 52 bodies.  (Six bodies have inadequate data to distinguish the mechanism or have very shallow surges.)  This is not to say that the SH mechanism is not operating (and indeed that it might dominate the slope of the phase curve for angles larger than observed), but that the measured slope of the phase curve over the small range of observable phase angles is dominated by the CB mechanism.
	
	(3) We present a graphical method for determining the Hapke surge parameters ($B_{C0}$ and $h_C$) based on the low-phase-angle slope of the phase curve ($S_0$) and the total drop in the phase curve between phase angles of $0\degr$ and $2\degr$ ($\Delta m_{0-2}$).  These two parameters are easy to read off a plotted phase curve.
	
	(4) We find that most of the KBOs and Centaurs have fairly linear phase curves over the observed range of phases, and this implies that typically the CB surge amplitude is relatively high (with $B_{C0} \gtrsim 0.5$) and that the surge width is relatively wide ($h_C \sim 3\degr$).  For moons, the surge widths can get rather narrow, with $h_c$ values from $0.1\degr$ to $0.3\degr$.
	
	(5) We find three correlations, where surge properties are significantly correlated with the color, the geometric albedo, and the inclination.  We attribute these relations to the various effects associated with the age of the surfaces, with young surfaces generally having low surge slopes, high albedos, and gray colors while old surfaces generally having steep surges, low albedos, and red colors.  We also find that no bodies have low surge slopes and low inclinations, so that low inclination bodies have steep surges, red colors, and a high frequency of binary companions.
	
	(6) Nereid has surface properties that are greatly different from those of KBOs and Centaurs, yet has surface properties similar to small inner icy moons.  Thus, we take this as a strong indication that Nereid is not a captured KBO but is instead a nearly-ejected inner Neptunian moon.

	The National Aeronautics and Space Administration provided funds under grants NAG5-13533 and NAG5-13369.

\clearpage

\begin{deluxetable}{lllllllll}
\tabletypesize{\scriptsize}
\tablecaption{Name, Orbit, Size, and Class
\label{tbl1}}
\tablewidth{0pt}
\tablehead{
\colhead{Name}   &
\colhead{Number}  &
\colhead{Designation}   &
\colhead{$i$ ($\degr$)}  &
\colhead{$e$}  &
\colhead{$a$ (AU)}  &
\colhead{$H$ (mag)}  &
\colhead{$D$ (km)}  &
\colhead{Class} 
}
\startdata
\underline{SMALL/RED:}	&		&		&		&		&		&		&		&		 \\
Elatus	 & 	31824	 & 	1999 UG5	 & 	5.6	 & 	0.418	 & 	12.8	 & 	9.9	 & 	30	 & 	Centaur (red)	 \\
\ldots	 & 	55638	 & 	2002 VE95	 & 	16.3	 & 	0.286	 & 	39.2	 & 	5.3	 & 	\ldots	 & 	Plutino (red)	 \\
Huya	 & 	38628	 & 	2000 EB173	 & 	15.5	 & 	0.275	 & 	39.4	 & 	4.7	 & 	533	 & 	Plutino (red)	 \\
Ixion	 & 	28978	 & 	2001 KX76	 & 	19.7	 & 	0.247	 & 	39.3	 & 	3.4	 & 	650	 & 	Plutino (red)	 \\
\ldots	 & 	119951	 & 	2002 KX14	 & 	0.4	 & 	0.043	 & 	38.8	 & 	4.5	 & 	\ldots	 & 	Plutino (red)	 \\
\ldots	 & 	47171	 & 	1999 TC36	 & 	8.4	 & 	0.229	 & 	39.6	 & 	4.7	 & 	415	 & 	Plutino (red)	 \\
\ldots	 & 	47932	 & 	2000 GN171	 & 	10.8	 & 	0.279	 & 	39.2	 & 	6.1	 & 	321	 & 	Plutino (red)	 \\
\ldots	 & 	120348	 & 	2004 TY364	 & 	24.8	 & 	0.064	 & 	38.8	 & 	4.5	 & 	\ldots	 & 	Plutino (red)	 \\
\ldots	 & 	55565	 & 	2002 AW197	 & 	24.3	 & 	0.128	 & 	47.5	 & 	3.3	 & 	735	 & 	Classical KBO (red)	 \\
\ldots	 & 	55637	 & 	2002 UX25	 & 	19.4	 & 	0.145	 & 	42.8	 & 	3.6	 & 	681	 & 	Classical KBO (red)	 \\
\ldots	 & 	40314	 & 	1999 KR16	 & 	24.9	 & 	0.299	 & 	48.5	 & 	5.7	 & 	\ldots	 & 	Classical KBO (red)	 \\
Varuna	 & 	20000	 & 	2000 WR106	 & 	17.1	 & 	0.056	 & 	43.2	 & 	3.5	 & 	500	 & 	Classical KBO (red)	 \\
\ldots	 & 	29981	 & 	1999 TD10	 & 	6.0	 & 	0.877	 & 	100.0	 & 	8.7	 & 	104	 & 	SDO (red)	 \\
\ldots	 & 	26375	 & 	1999 DE9	 & 	7.6	 & 	0.423	 & 	55.9	 & 	4.8	 & 	461	 & 	SDO (red)	 \\
\underline{SMALL/GRAY:}	&		&		&		&		&		&		&		&		 \\
\ldots	 & 	95626	 & 	2002 GZ32	 & 	15.0	 & 	0.222	 & 	23.2	 & 	6.8	 & 	\ldots	 & 	Centaur (gray)	 \\
\ldots	 & 	73480	 & 	2002 PN34	 & 	16.6	 & 	0.572	 & 	31.1	 & 	8.2	 & 	120	 & 	Centaur (gray)	 \\
Asbolus	 & 	8405	 & 	1995 GO	 & 	17.6	 & 	0.618	 & 	17.9	 & 	9.1	 & 	84	 & 	Centaur (gray)	 \\
Thereus	 & 	32532	 & 	2001 PT13	 & 	20.4	 & 	0.197	 & 	10.6	 & 	8.9	 & 	88	 & 	Centaur (gray)	 \\
Bienor	 & 	54598	 & 	2000 QC243	 & 	20.8	 & 	0.201	 & 	16.5	 & 	7.6	 & 	207	 & 	Centaur (gray)	 \\
Hylonome	 & 	10370	 & 	1995 DW2	 & 	4.1	 & 	0.245	 & 	25.0	 & 	8.4	 & 	70	 & 	Centaur (gray)	 \\
Typhon	 & 	42355	 & 	2002 CR46	 & 	2.4	 & 	0.543	 & 	38.4	 & 	7.2	 & 	175	 & 	Centaur (gray)	 \\
Pelion	 & 	49036	 & 	1998 QM107	 & 	9.4	 & 	0.138	 & 	20.0	 & 	10.4	 & 	\ldots	 & 	Centaur (gray)	 \\
Echeclus	 & 	60558	 & 	2000 EC98	 & 	4.3	 & 	0.455	 & 	10.8	 & 	8.6	 & 	84	 & 	Centaur (gray, active)	 \\
\ldots	 & 	145451	 & 	2005 RM43	 & 	28.8	 & 	0.610	 & 	90.0	 & 	4.4	 & 	\ldots	 & 	SDO (gray)	 \\
\underline{INTERMEDIATE:}	&		&		&		&		&		&		&		&		 \\
Charon	 & 	\ldots	 & 	PI	 & 	17.2	 & 	0.250	 & 	39.4	 & 	0.9	 & 	1186	 & 	Moon (Pluto) \& Plutino (gray)	 \\
Orcus	 & 	90482	 & 	2004 DW	 & 	20.5	 & 	0.222	 & 	39.4	 & 	2.3	 & 	946	 & 	Plutino (gray)	 \\
Quaoar	 & 	50000	 & 	2002 LM60	 & 	8.0	 & 	0.037	 & 	43.1	 & 	2.7	 & 	844	 & 	Classical KBO (red)	 \\
\underline{LARGE:}	&		&		&		&		&		&		&		&		 \\
Pluto	 & 	134340	 & 	\ldots	 & 	17.2	 & 	0.250	 & 	39.4	 & 	-0.8	 & 	2390	 & 	Dwarf Planet \& Plutino	 \\
Makemake	 & 	136472	 & 	2005 FY9	 & 	29.0	 & 	0.156	 & 	45.7	 & 	-0.5	 & 	1500	 & 	Dwarf Planet \& Classical KBO	 \\
Eris	 & 	136199	 & 	2003 UB313	 & 	44.1	 & 	0.441	 & 	67.7	 & 	-1.2	 & 	2600	 & 	Dwarf Planet \& SDO	 \\
Sedna	 & 	90377	 & 	2003 VB122	 & 	11.9	 & 	0.854	 & 	520.4	 & 	1.5	 & 	$<$1600	 & 	SDO or Inner Oort Cloud	 \\
Triton	 & 	\ldots	 & 	NI	 & 	1.8	 & 	0.009	 & 	30.1	 & 	-1.2	 & 	2706	 & 	Captured KBO \& Moon (Neptune)	 \\
\underline{COLLISIONAL:}	&		&		&		&		&		&		&		&		 \\
Haumea	 & 	136108	 & 	2003 EL61	 & 	28.2	 & 	0.189	 & 	43.3	 & 	0.2	 & 	1150	 & 	EL61 family	 \\
\ldots	 & 	145453	 & 	2005 RR43	 & 	28.6	 & 	0.138	 & 	43.1	 & 	4.0	 & 	\ldots	 & 	EL61 family	 \\
\ldots	 & 	24835	 & 	1995 SM55	 & 	27.0	 & 	0.108	 & 	42.1	 & 	4.7	 & 	\ldots	 & 	EL61 family	 \\
\ldots	 & 	120178	 & 	2003 OP32	 & 	27.2	 & 	0.109	 & 	43.4	 & 	4.1	 & 	\ldots	 & 	EL61 family	 \\
\ldots	 & 	55636	 & 	2002 TX300	 & 	25.9	 & 	0.124	 & 	43.5	 & 	3.1	 & 	$<$800	 & 	EL61 family	 \\
\underline{MOONS:}	&		&		&		&		&		&		&		&		 \\
Europa	 & 	\ldots	 & 	JII	 & 	1.3	 & 	0.048	 & 	5.2	 & 	-1.4	 & 	3138	 & 	Moon (Jupiter)	 \\
Ganymede	 & 	\ldots	 & 	JIII	 & 	1.3	 & 	0.048	 & 	5.2	 & 	-2.1	 & 	5262	 & 	Moon (Jupiter)	 \\
Callisto	 & 	\ldots	 & 	JIV	 & 	1.3	 & 	0.048	 & 	5.2	 & 	-1.1	 & 	4800	 & 	Moon (Jupiter)	 \\
Enceladus	 & 	\ldots	 & 	SII	 & 	2.5	 & 	0.054	 & 	9.5	 & 	2.1	 & 	500	 & 	Moon (Saturn)	 \\
Tethys	 & 	\ldots	 & 	SIII	 & 	2.5	 & 	0.054	 & 	9.5	 & 	0.6	 & 	1060	 & 	Moon (Saturn)	 \\
Dione	 & 	\ldots	 & 	SIV	 & 	2.5	 & 	0.054	 & 	9.5	 & 	0.8	 & 	1120	 & 	Moon (Saturn)	 \\
Rhea	 & 	\ldots	 & 	SV	 & 	2.5	 & 	0.054	 & 	9.5	 & 	0.1	 & 	1530	 & 	Moon (Saturn)	 \\
Iapetus	 & 	\ldots	 & 	SVIII	 & 	2.5	 & 	0.054	 & 	9.5	 & 	1.5	 & 	1460	 & 	Moon (Saturn)	 \\
Ariel	 & 	\ldots	 & 	UI	 & 	0.8	 & 	0.047	 & 	19.2	 & 	1.4	 & 	1158	 & 	Moon (Uranus)	 \\
Umbriel	 & 	\ldots	 & 	UII	 & 	0.8	 & 	0.047	 & 	19.2	 & 	2.1	 & 	1172	 & 	Moon (Uranus)	 \\
Titania	 & 	\ldots	 & 	UIII	 & 	0.8	 & 	0.047	 & 	19.2	 & 	1.0	 & 	1580	 & 	Moon (Uranus)	 \\
Oberon	 & 	\ldots	 & 	UIV	 & 	0.8	 & 	0.047	 & 	19.2	 & 	1.2	 & 	1524	 & 	Moon (Uranus)	 \\
Miranda	 & 	\ldots	 & 	UV	 & 	0.8	 & 	0.047	 & 	19.2	 & 	3.6	 & 	480	 & 	Moon (Uranus)	 \\
Puck	 & 	\ldots	 & 	UXV	 & 	0.8	 & 	0.047	 & 	19.2	 & 	7.5	 & 	154	 & 	Moon (Uranus)	 \\
Nereid	 & 	\ldots	 & 	NII	 & 	1.8	 & 	0.009	 & 	30.1	 & 	4.4	 & 	340	 & 	Moon (Neptune)	 \\
\enddata
    
\end{deluxetable}

\clearpage

\begin{deluxetable}{llllll}
\tabletypesize{\scriptsize}
\tablecaption{Albedo, Color, and Composition
\label{tbl2}}
\tablewidth{0pt}
\tablehead{
\colhead{Object}   &
\colhead{$p_V$ (\%)}  &
\colhead{$B-I$ (mag)}   &
\colhead{Ref.\tablenotemark{a}}  &
\colhead{Composition}  &
\colhead{Ref.\tablenotemark{a}} 
}
\startdata
\underline{SMALL/RED:}	 & 		 & 		 & 		 & 		 & 		 \\
Elatus	 & 	10	 & 	2.20	 & 	2	 & 	H$_2$O+Tholin	 & 	14	 \\
2002 VE95	 & 	\ldots	 & 	2.53	 & 	RST2007	 & 	H$_2$O+CH$_3$OH	 & 	15	 \\
Huya	 & 	5	 & 	1.97	 & 	RST2007	 & 	\ldots	 & 	\ldots	 \\
Ixion	 & 	12	 & 	2.02	 & 	RST2007	 & 	H$_2$O+Tholin	 & 	16	 \\
2002 KX14	 & 	\ldots	 & 	2.3	 & 	3	 & 	\ldots	 & 	\ldots	 \\
1999 TC36	 & 	7	 & 	2.48	 & 	RST2007	 & 	H$_2$O+Tholin	 & 	17	 \\
2000 GN171	 & 	6	 & 	2.16	 & 	4	 & 	\ldots	 & 	\ldots	 \\
2004 TY364	 & 	\ldots	 & 	2.34	 & 	RST2007	 & 	\ldots	 & 	\ldots	 \\
2002 AW197	 & 	12	 & 	2.13	 & 	RST2007	 & 	H$_2$O+Tholin	 & 	18	 \\
2002 UX25	 & 	11	 & 	2.07	 & 	RST2007	 & 	\ldots	 & 	\ldots	 \\
1999 KR16	 & 	\ldots	 & 	2.61	 & 	5	 & 	\ldots	 & 	\ldots	 \\
Varuna	 & 	16	 & 	2.17	 & 	RST2007	 & 	H$_2$O	 & 	19	 \\
1999 TD10	 & 	4	 & 	2.01	 & 	RST2007	 & 	\ldots	 & 	\ldots	 \\
1999 DE9	 & 	7	 & 	2.12	 & 	RST2007	 & 	H$_2$O+Tholin	 & 	20	 \\
\underline{SMALL/GRAY:}	 & 		 & 		 & 		 & 		 & 		 \\
2002 GZ32	 & 	\ldots	 & 	1.81	 & 	RST2007	 & 	\ldots	 & 	\ldots	 \\
2002 PN34	 & 	4	 & 	1.95	 & 	RST2007	 & 	\ldots	 & 	\ldots	 \\
Asbolus	 & 	5	 & 	1.69	 & 	RST2007	 & 	H$_2$O [patchy]	 & 	21	 \\
Thereus	 & 	4	 & 	1.76	 & 	RST2007	 & 	H$_2$O [patchy]	 & 	22	 \\
Bienor	 & 	3	 & 	1.70	 & 	RST2007	 & 	kerogen+ 3\%H$_2$O+olivine	 & 	17	 \\
Hylonome	 & 	6	 & 	1.68	 & 	6	 & 	\ldots	 & 	\ldots	 \\
Typhon	 & 	5	 & 	1.76	 & 	RST2007	 & 	\ldots	 & 	\ldots	 \\
Pelion	 & 	\ldots	 & 	1.61	 & 	4	 & 	\ldots	 & 	\ldots	 \\
Echeclus	 & 	4	 & 	1.76	 & 	4	 & 	\ldots	 & 	\ldots	 \\
2005 RM43	 & 	\ldots	 & 	1.32	 & 	RST2007	 & 	\ldots	 & 	\ldots	 \\
\underline{INTERMEDIATE:}	 & 		 & 		 & 		 & 		 & 		 \\
Charon	 & 	38	 & 	1.47	 & 	7	 & 	crystalline H$_2$O+NH$_3$H$_2$O	 & 	23	 \\
Orcus	 & 	20	 & 	1.45	 & 	RST2007	 & 	crystalline H$_2$O+NH$_3$+NH$_3$H$_2$O	 & 	24	 \\
Quaoar	 & 	20	 & 	2.47	 & 	RST2007	 & 	crystalline H$_2$O+NH$_3$+C$_2$H$_6$	 & 	25	 \\
\underline{LARGE:}	 & 		 & 		 & 		 & 		 & 		 \\
Pluto	 & 	61	 & 	1.74	 & 	5	 & 	CH$_4$+N$_2$	 & 	26	 \\
Makemake	 & 	80	 & 	1.64	 & 	RST2007	 & 	CH$_4$+C$_2$H$_6$	 & 	27	 \\
Eris	 & 	70	 & 	1.60	 & 	RST2007	 & 	CH$_4$(+N$_2$)	 & 	28	 \\
Sedna	 & 	$>$16	 & 	2.62	 & 	RST2007	 & 	CH$_4$+N$_2$(+H$_2$O)	 & 	29	 \\
Triton	 & 	70	 & 	1.82	 & 	8	 & 	CH$_4$+N$_2$	 & 	26	 \\
\underline{COLLISIONAL:}	 & 		 & 		 & 		 & 		 & 		 \\
Haumea	 & 	84	 & 	1.33	 & 	RST2007	 & 	crystalline H$_2$O	 & 	30	 \\
2005 RR43	 & 	\ldots	 & 	1.49	 & 	This paper	 & 	crystalline H$_2$O	 & 	31	 \\
1995 SM55	 & 	\ldots	 & 	1.35	 & 	This paper	 & 	crystalline H$_2$O	 & 	32	 \\
2003 OP32	 & 	\ldots	 & 	1.45	 & 	This paper	 & 	crystalline H$_2$O	 & 	33	 \\
2002 TX300	 & 	$>$19	 & 	1.51	 & 	RST2007	 & 	crystalline H$_2$O	 & 	34	 \\
\underline{MOONS:}	 & 		 & 		 & 		 & 		 & 		 \\
Europa	 & 	64	 & 	1.59	 & 	9	 & 	crystalline H$_2$O	 & 	35	 \\
Ganymede	 & 	42	 & 	1.68	 & 	9	 & 	crystalline H$_2$O	 & 	35	 \\
Callisto	 & 	20	 & 	1.78	 & 	9	 & 	crystalline H$_2$O	 & 	35	 \\
Enceladus	 & 	100	 & 	1.25	 & 	10	 & 	crystalline H$_2$O	 & 	36	 \\
Tethys	 & 	90	 & 	1.96	 & 	11	 & 	crystalline H$_2$O	 & 	36	 \\
Dione	 & 	70	 & 	1.92	 & 	11	 & 	crystalline H$_2$O	 & 	36	 \\
Rhea	 & 	70	 & 	2.00	 & 	11	 & 	crystalline H$_2$O	 & 	36	 \\
Iapetus	 & 	50	 & 	1.98	 & 	11	 & 	crystalline H$_2$O	 & 	36	 \\
Ariel	 & 	35	 & 	1.53	 & 	12	 & 	crystalline H$_2$O	 & 	12, 37	 \\
Umbriel	 & 	19	 & 	1.50	 & 	12	 & 	crystalline H$_2$O	 & 	12, 37	 \\
Titania	 & 	23	 & 	1.61	 & 	12	 & 	crystalline H$_2$O	 & 	12, 37	 \\
Oberon	 & 	21	 & 	1.60	 & 	12	 & 	crystalline H$_2$O	 & 	12, 37	 \\
Miranda	 & 	31	 & 	1.44	 & 	12	 & 	H$_2$O	 & 	12	 \\
Puck	 & 	9	 & 	1.51	 & 	12	 & 	H$_2$O	 & 	12	 \\
Nereid	 & 	40	 & 	1.43	 & 	13	 & 	crystalline H$_2$O	 & 	38	 \\
\enddata
    
\tablenotetext{a}{1	Rabinowitz, Schaefer, \& Tourtellotte 2007 (RST2007).
2	Peixinho et al. 2001.
3	Tegler \& Romanishin 2007.
4	Boehnhardt et al. 2002.
5	Jewitt \& Luu 2001.
6	Doressoundiram et al. 2002.
7	Fink \& Disanti 1988.
8	Cruikshank \& Brown 1986.
9	McFadden, Bell, \& McCord 1980.
10	Verbiscer, French, \& McGhee 2005.
11	Noland et al. 1974.
12	Karkoschka 2001.
13	Schaefer \& Schaefer 2000.
14	Bauer et al. 2002.
15	Barucci et al. 2006.
16	Boehnhardt et al. 2004.
17	Dotto et al. 2003.
18	Doressoundiram et al. 2005.
19	Licandro, Oliva, \& Di Martino 2001.
20	Doressoundiram et al. 2003.
21	Kern et al. 2000.
22	Licandro \& Pinilla-Alonso 2005.
23	Cook et al. 2007.
24	Cook, Desch, \& Roush 2007.
25	Jewitt \& Luu 2004, Schaller \& Brown 2007b.
26	Grundy, Buie, \& Spencer 2002.
27	Brown et al. 2007a.
28	Licandro et al. 2006a.
29	Emery et al. 2007.
30	Trujillo et al. 2007.
31	Pinilla-Alonso et al. 2007.
32	Barkume, Brown, \& Schaller 2008.
33	Brown et al. 2007b.
34	Licandro et al. 2006b.
35	Johnson \& Pilcher 1977.
36	Clark, Fanale, \& Gaffey 1986.
37	Grundy et al. 1999.
38	Brown 2000.}

\end{deluxetable}

\clearpage

\begin{deluxetable}{llllll}
\tabletypesize{\scriptsize}
\tablecaption{Surge Properties
\label{tbl3}}
\tablewidth{0pt}
\tablehead{
\colhead{Object}   &
\colhead{$S_0$ (mag deg$^{-1}$)}  &
\colhead{$S_1$ (mag deg$^{-1}$)}  &
\colhead{$\Delta m_{0-2}$(mag)}   &
\colhead{$S_B-S_I$}  &
\colhead{Ref.\tablenotemark{a}} 
}
\startdata

\underline{SMALL/RED:}	 & 				 & 				 & 				 & 				 & 		 \\
Elatus	 & 	\ldots			 & 	0.18	 $\pm$ 	0.05	 & 	\ldots			 & 	0.00	$\pm$	0.02	 & 	 1-6	 \\
2002 VE95	 & 	\ldots			 & 	0.11	 $\pm$ 	0.02	 & 	0.22	$\pm$	0.04	 & 	0.02	$\pm$	0.04	 & 	RST2007	 \\
Huya	 & 	0.13	 $\pm$ 	0.01	 & 	0.13	 $\pm$ 	0.01	 & 	0.26	$\pm$	0.02	 & 	$\sim$0.00			 & 	RST2007	 \\
Ixion	 & 	0.15	 $\pm$ 	0.03	 & 	0.15	 $\pm$ 	0.03	 & 	0.20	-	0.30	 & 	0.12	$\pm$	0.08	 & 	RST2007	 \\
2002 KX14	 & 	0.18	 $\pm$ 	0.03	 & 	0.18	 $\pm$ 	0.03	 & 	0.25	-	0.36	 & 	\ldots			 & 	RST2007	 \\
1999 TC36	 & 	0.21	 $\pm$ 	0.03	 & 	0.21	 $\pm$ 	0.03	 & 	0.42	$\pm$	0.15	 & 	0.02	$\pm$	0.07	 & 	RST2007	 \\
2000 GN171	 & 	0.21	 $\pm$ 	0.02	 & 	0.21	 $\pm$ 	0.02	 & 	0.42	$\pm$	0.04	 & 	$<$-0.14	$\pm$	0.05	 & 	RST2007	 \\
2004 TY364	 & 	\ldots			 & 	0.22	 $\pm$ 	0.03	 & 	\ldots			 & 	-0.28	$\pm$	0.08	 & 	RST2007	 \\
2002 AW197	 & 	0.10	 $\pm$ 	0.03	 & 	0.10	 $\pm$ 	0.03	 & 	0.12	-	0.20	 & 	-0.08	$\pm$	0.07	 & 	RST2007	 \\
2002 UX25	 & 	0.14	 $\pm$ 	0.02	 & 	0.14	 $\pm$ 	0.02	 & 	0.19	-	0.28	 & 	0.01	$\pm$	0.04	 & 	RST2007	 \\
1999 KR16	 & 	0.28	 $\pm$ 	0.05	 & 	0.12	 $\pm$ 	0.03	 & 	0.30	$\pm$	0.05	 & 	$\sim$0.00			 & 	7	 \\
Varuna	 & 	0.25	 $\pm$ 	0.02	 & 	0.25	 $\pm$ 	0.02	 & 	0.34	-	0.50	 & 	0.06	$\pm$	0.05	 & 	RST2007	 \\
1999 TD10	 & 	0.12	 $\pm$ 	0.01	 & 	0.12	 $\pm$ 	0.01	 & 	0.24	$\pm$	0.10	 & 	-0.05	$\pm$	0.02	 & 	RST2007	 \\
1999 DE9	 & 	0.16	 $\pm$ 	0.02	 & 	0.16	 $\pm$ 	0.02	 & 	0.32	$\pm$	0.04	 & 	0.02	$\pm$	0.04	 & 	RST2007	 \\
\underline{SMALL/GRAY:}	 & 				 & 				 & 				 & 				 & 		 \\
2002 GZ32	 & 	\ldots			 & 	0.01	 $\pm$ 	0.03	 & 	0.02	$\pm$	0.10	 & 	0.03	$\pm$	0.13	 & 	RST2007	 \\
2002 PN34	 & 	0.04	 $\pm$ 	0.01	 & 	0.04	 $\pm$ 	0.01	 & 	0.08	$\pm$	0.10	 & 	-0.02	$\pm$	0.01	 & 	RST2007	 \\
Asbolus	 & 	\ldots			 & 	0.05	 $\pm$ 	0.01	 & 	\ldots			 & 	0.00	$\pm$	0.01	 & 	RST2007	 \\
Thereus	 & 	\ldots			 & 	0.07	 $\pm$ 	0.01	 & 	0.14	$\pm$	0.06	 & 	0.01	$\pm$	0.01	 & 	RST2007	 \\
Bienor	 & 	\ldots			 & 	0.10	 $\pm$ 	0.01	 & 	0.20	$\pm$	0.10	 & 	-0.03	$\pm$	0.02	 & 	RST2007	 \\
Hylonome	 & 	0.13	 $\pm$ 	0.06	 & 	0.13	 $\pm$ 	0.06	 & 	0.18	$\pm$	0.05	 & 	0.00	$\pm$	0.05	 & 	1, 2, 8-12	 \\
Typhon	 & 	0.13	 $\pm$ 	0.01	 & 	0.13	 $\pm$ 	0.01	 & 	0.26	$\pm$	0.02	 & 	0.00	$\pm$	0.03	 & 	RST2007	 \\
Pelion	 & 	0.7	 $\pm$ 	0.3	 & 	0.11	 $\pm$ 	0.03	 & 	0.33	$\pm$	0.10	 & 	\ldots			 & 	9	 \\
Echeclus	 & 	0.18	 $\pm$ 	0.02	 & 	0.18	 $\pm$ 	0.02	 & 	0.36	$\pm$	0.04	 & 	$\sim$0.00			 & 	13	 \\
2005 RM43	 & 	0.18	 $\pm$ 	0.03	 & 	0.18	 $\pm$ 	0.03	 & 	\ldots			 & 	0.00	$\pm$	0.05	 & 	RSST2008	 \\
\underline{INTERMEDIATE:}	 & 				 & 				 & 				 & 				 & 		 \\
Charon	 & 	\ldots			 & 	0.09	 $\pm$ 	0.01	 & 	0.17	$\pm$	0.02	 & 	$\sim$0.00			 & 	14	 \\
Orcus	 & 	\ldots			 & 	0.13	 $\pm$ 	0.02	 & 	\ldots			 & 	-0.09	$\pm$	0.04	 & 	RST2007	 \\
Quaoar	 & 	0.14	 $\pm$ 	0.02	 & 	0.14	 $\pm$ 	0.02	 & 	0.19	-	0.28	 & 	-0.22	$\pm$	0.05	 & 	RST2007	 \\
\underline{LARGE:}	 & 				 & 				 & 				 & 				 & 		 \\
Pluto	 & 	\ldots			 & 	0.03	 $\pm$ 	0.01	 & 	0.06	$\pm$	0.002	 & 	$\sim$0.00			 & 	14	 \\
Makemake	 & 	\ldots			 & 	0.06	 $\pm$ 	0.01	 & 	\ldots			 & 	0.02	$\pm$	0.03	 & 	RST2007	 \\
Eris	 & 	0.08	 $\pm$ 	0.01	 & 	\ldots			 & 	\ldots			 & 	-0.14	$\pm$	0.04	 & 	RST2007	 \\
Sedna	 & 	0.15	 $\pm$ 	0.03	 & 	\ldots			 & 	\ldots			 & 	\ldots			 & 	RST2007	 \\
Triton	 & 	0.92	 $\pm$ 	0.18	 & 	0.00	 $\pm$ 	0.01	 & 	0.10	$\pm$	0.02	 & 	$\sim$0.00			 & 	15	 \\
\underline{COLLISIONAL:}	 & 				 & 				 & 				 & 				 & 		 \\
Haumea	 & 	\ldots			 & 	0.11	 $\pm$ 	0.01	 & 	\ldots			 & 	-0.04	$\pm$	0.02	 & 	RST2007	 \\
2005 RR43	 & 	\ldots			 & 	0.00	 $\pm$ 	0.03	 & 	\ldots			 & 	-0.13	$\pm$	0.04	 & 	RSST2008	 \\
1995 SM55	 & 	0.06	 $\pm$ 	0.04	 & 	0.06	 $\pm$ 	0.04	 & 	\ldots			 & 	0.00	$\pm$	0.09	 & 	RSST2008	 \\
2003 OP32	 & 	\ldots			 & 	0.01	 $\pm$ 	0.04	 & 	\ldots			 & 	-0.03	$\pm$	0.06	 & 	RSST2008	 \\
2002 TX300	 & 	\ldots			 & 	0.05	 $\pm$ 	0.03	 & 	\ldots			 & 	-0.10	$\pm$	0.07	 & 	RST2007	 \\
\underline{MOONS:}	 & 				 & 				 & 				 & 				 & 		 \\
Europa	 & 	0.24	 $\pm$ 	0.08	 & 	0.04	 $\pm$ 	0.02	 & 	0.15	$\pm$	0.03	 & 	$\sim$0.00			 & 	16, 17	 \\
Ganymede	 & 	0.04	 $\pm$ 	0.01	 & 	0.04	 $\pm$ 	0.01	 & 	0.07	$\pm$	0.01	 & 	$\sim$0.00			 & 	16, 18	 \\
Callisto	 & 	0.08	 $\pm$ 	0.02	 & 	0.07	 $\pm$ 	0.01	 & 	0.15	$\pm$	0.04	 & 	$\sim$0.00			 & 	16, 18	 \\
Enceladus	 & 	\ldots			 & 	0.36	 $\pm$ 	0.16	 & 	0.25	$\pm$	0.07	 & 	0.02	$\pm$	0.02	 & 	19	 \\
Tethys	 & 	0.02	 $\pm$ 	0.01	 & 	0.02	 $\pm$ 	0.01	 & 	0.00	$\pm$	0.03	 & 	0.00	$\pm$	0.02	 & 	20	 \\
Dione	 & 	0.01	 $\pm$ 	0.01	 & 	0.01	 $\pm$ 	0.01	 & 	0.00	$\pm$	0.03	 & 	0.03	$\pm$	0.01	 & 	20	 \\
Rhea	 & 	0.16	 $\pm$ 	0.06	 & 	0.06	 $\pm$ 	0.02	 & 	0.15	$\pm$	0.03	 & 	0.00	$\pm$	0.01	 & 	20, 21	 \\
Iapetus	 & 	0.17	 $\pm$ 	0.04	 & 	0.17	 $\pm$ 	0.04	 & 	0.20	$\pm$	0.03	 & 	0.01	$\pm$	0.01	 & 	20, 22	 \\
Ariel	 & 	1.2	 $\pm$ 	0.2	 & 	0.03	 $\pm$ 	0.01	 & 	0.50	$\pm$	0.02	 & 	0.03	$\pm$	0.10	 & 	23	 \\
Umbriel	 & 	0.5	 $\pm$ 	0.1	 & 	0.07	 $\pm$ 	0.01	 & 	0.42	$\pm$	0.02	 & 	0.01	$\pm$	0.10	 & 	23	 \\
Titania	 & 	0.7	 $\pm$ 	0.1	 & 	0.07	 $\pm$ 	0.01	 & 	0.45	$\pm$	0.02	 & 	0.01	$\pm$	0.10	 & 	23	 \\
Oberon	 & 	0.9	 $\pm$ 	0.1	 & 	0.07	 $\pm$ 	0.01	 & 	0.44	$\pm$	0.02	 & 	-0.10	$\pm$	0.10	 & 	23	 \\
Miranda	 & 	1.0	 $\pm$ 	0.2	 & 	0.05	 $\pm$ 	0.01	 & 	0.49	$\pm$	0.02	 & 	-0.03	$\pm$	0.12	 & 	23	 \\
Puck	 & 	1.0	 $\pm$ 	0.2	 & 	0.08	 $\pm$ 	0.02	 & 	0.35	$\pm$	0.06	 & 	\ldots			 & 	23	 \\
Nereid	 & 	0.50	 $\pm$ 	0.05	 & 	0.15	 $\pm$ 	0.03	 & 	0.41	$\pm$	0.02	 & 	0.10	$\pm$	0.04	 & 	RST2007	 \\
\enddata
    
\tablenotetext{a}{RST2007 is Rabinowitz, Schaefer, \& Tourtellotte 2007.  
         RSST2008 is Rabinowitz et al. 2008.
1	Doressoundiram et al. 2002.
2	Tegler \& Romanishin 2003.
3	Delsanti et al. 2001.
4	Gutierrez et al. 2001.
5	Peixinho et al. 2001.
6	Bauer et al. 2002.
7	Sheppard \& Jewitt 2002.
8	Tegler \& Romanishin 1998.
9	Bauer et al 2003
10	Luu \& Jewitt 1996.
11	Green et al. 1997.
12	Magnusson et al. 1998.
13	Rousselot et al. 2005.
14	Buie, Tholen, \& Wasserman 1997.
15	Buratti, Bauer, \& Hicks 2007.
16	Domingue \& Verbiscer 1997.
17	Domingue et al. 1991.
18	Millis \& Thompson 1975.
19	Verbiscer, French, \& McGhee 2005.
20	Noland et al. 1974.
21	Domingue, Lockwood, \& Thompson 1995.
22	Franklin \& Cook 1974.
23	Karkoschka 2001.}
    
\end{deluxetable}

\clearpage

\begin{deluxetable}{llllll}
\tabletypesize{\scriptsize}
\tablecaption{Dominant Surge Mechanism
\label{tbl4}}
\tablewidth{0pt}
\tablehead{
\colhead{Object}   &
\colhead{Color Independence}  &
\colhead{Surge Slope}   &
\colhead{Phase Curve Shape}  &
\colhead{Albedo}  &
\colhead{Dominant Mechanism} 
}
\startdata

\underline{SMALL/RED:}	 & 		 & 		 & 		 & 		 & 		 \\
Elatus	 & 	SH or CB	 & 	CB	 & 	\ldots	 & 	SH or CB	 & 	CB	 \\
2002 VE95	 & 	SH or CB	 & 	CB	 & 	\ldots	 & 	\ldots	 & 	CB	 \\
Huya	 & 	SH or CB	 & 	CB	 & 	\ldots	 & 	SH or CB	 & 	CB	 \\
Ixion	 & 	SH or CB	 & 	CB	 & 	\ldots	 & 	SH or CB	 & 	CB	 \\
2002 KX14	 & 	\ldots	 & 	CB	 & 	\ldots	 & 	\ldots	 & 	CB	 \\
1999 TC36	 & 	SH or CB	 & 	CB	 & 	\ldots	 & 	SH or CB	 & 	CB	 \\
2000 GN171	 & 	CB	 & 	CB	 & 	\ldots	 & 	SH or CB	 & 	CB	 \\
2004 TY364	 & 	CB	 & 	CB	 & 	\ldots	 & 	\ldots	 & 	CB	 \\
2002 AW197	 & 	SH or CB	 & 	CB	 & 	\ldots	 & 	SH or CB	 & 	CB	 \\
2002 UX25	 & 	SH or CB	 & 	CB	 & 	\ldots	 & 	SH or CB	 & 	CB	 \\
1999 KR16	 & 	SH or CB	 & 	CB	 & 	\ldots	 & 	\ldots	 & 	CB	 \\
Varuna	 & 	SH or CB	 & 	CB	 & 	\ldots	 & 	SH or CB	 & 	CB	 \\
1999 TD10	 & 	?	 & 	CB	 & 	\ldots	 & 	SH or CB	 & 	CB	 \\
1999 DE9	 & 	SH or CB	 & 	CB	 & 	\ldots	 & 	\ldots	 & 	CB	 \\
\underline{SMALL/GRAY:}	 & 		 & 		 & 		 & 		 & 		 \\
2002 GZ32	 & 	SH or CB	 & 	SH or CB	 & 	\ldots	 & 	\ldots	 & 	no surge	 \\
2002 PN34	 & 	?	 & 	SH or CB	 & 	\ldots	 & 	SH or CB	 & 	?	 \\
Asbolus	 & 	SH or CB	 & 	SH or CB	 & 	\ldots	 & 	SH or CB	 & 	?	 \\
Thereus	 & 	SH or CB	 & 	CB	 & 	\ldots	 & 	SH or CB	 & 	CB	 \\
Bienor	 & 	SH or CB	 & 	CB	 & 	\ldots	 & 	SH or CB	 & 	CB	 \\
Hylonome	 & 	SH or CB	 & 	CB	 & 	\ldots	 & 	SH or CB	 & 	CB	 \\
Typhon	 & 	SH or CB	 & 	CB	 & 	\ldots	 & 	SH or CB	 & 	CB	 \\
Pelion	 & 	\ldots	 & 	CB	 & 	\ldots	 & 	\ldots	 & 	CB	 \\
Echeclus	 & 	SH or CB	 & 	CB	 & 	\ldots	 & 	SH or CB	 & 	CB	 \\
2005 RM43	 & 	SH or CB	 & 	CB	 & 	\ldots	 & 	\ldots	 & 	CB	 \\
\underline{INTERMEDIATE:}	 & 		 & 		 & 		 & 		 & 		 \\
Charon	 & 	SH or CB	 & 	CB	 & 	\ldots	 & 	SH or CB	 & 	CB	 \\
Orcus	 & 	CB	 & 	CB	 & 	\ldots	 & 	SH or CB	 & 	CB	 \\
Quaoar	 & 	CB	 & 	CB	 & 	\ldots	 & 	SH or CB	 & 	CB	 \\
\underline{LARGE:}	 & 		 & 		 & 		 & 		 & 		 \\
Pluto	 & 	SH or CB	 & 	SH or CB	 & 	\ldots	 & 	CB	 & 	CB	 \\
Makemake	 & 	SH or CB	 & 	SH or CB	 & 	\ldots	 & 	CB	 & 	CB	 \\
Eris	 & 	CB	 & 	CB	 & 	\ldots	 & 	CB	 & 	CB	 \\
Sedna	 & 	\ldots	 & 	CB	 & 	\ldots	 & 	\ldots	 & 	CB	 \\
Triton	 & 	SH or CB	 & 	CB	 & 	\ldots	 & 	CB	 & 	CB	 \\
\underline{COLLISIONAL:}	 & 		 & 		 & 		 & 		 & 		 \\
Haumea	 & 	?	 & 	CB	 & 	\ldots	 & 	CB	 & 	CB	 \\
2005 RR43	 & 	CB	 & 	SH or CB	 & 	\ldots	 & 	\ldots	 & 	CB	 \\
1995 SM55	 & 	SH or CB	 & 	SH or CB	 & 	\ldots	 & 	\ldots	 & 	?	 \\
2003 OP32	 & 	SH or CB	 & 	SH or CB	 & 	\ldots	 & 	\ldots	 & 	no surge	 \\
2002 TX300	 & 	SH or CB	 & 	SH or CB	 & 	\ldots	 & 	\ldots	 & 	?	 \\
\underline{MOONS:}	 & 		 & 		 & 		 & 		 & 		 \\
Europa	 & 	SH or CB	 & 	CB	 & 	CB	 & 	CB	 & 	CB	 \\
Ganymede	 & 	SH or CB	 & 	CB	 & 	\ldots	 & 	CB	 & 	CB	 \\
Callisto	 & 	SH or CB	 & 	CB	 & 	\ldots	 & 	SH or CB	 & 	CB	 \\
Enceladus	 & 	SH or CB	 & 	CB	 & 	CB	 & 	CB	 & 	CB	 \\
Tethys	 & 	SH or CB	 & 	SH or CB	 & 	\ldots	 & 	CB	 & 	CB	 \\
Dione	 & 	SH or CB	 & 	SH or CB	 & 	\ldots	 & 	CB	 & 	CB	 \\
Rhea	 & 	SH or CB	 & 	CB	 & 	\ldots	 & 	CB	 & 	CB	 \\
Iapetus	 & 	SH or CB	 & 	CB	 & 	\ldots	 & 	CB	 & 	CB	 \\
Ariel	 & 	SH or CB	 & 	CB	 & 	CB	 & 	SH or CB	 & 	CB	 \\
Umbriel	 & 	SH or CB	 & 	CB	 & 	CB	 & 	SH or CB	 & 	CB	 \\
Titania	 & 	SH or CB	 & 	CB	 & 	CB	 & 	SH or CB	 & 	CB	 \\
Oberon	 & 	SH or CB	 & 	CB	 & 	CB	 & 	SH or CB	 & 	CB	 \\
Miranda	 & 	SH or CB	 & 	CB	 & 	CB	 & 	SH or CB	 & 	CB	 \\
Puck	 & 	\ldots	 & 	CB	 & 	CB	 & 	SH or CB	 & 	CB	 \\
Nereid	 & 	CB	 & 	CB	 & 	CB	 & 	SH or CB	 & 	CB	 \\

\enddata    
    
\end{deluxetable}

\clearpage

\begin{deluxetable}{llllll}
\tabletypesize{\scriptsize}
\tablecaption{Model fits to Nereid's Phase Curve 
\label{tbl5}}
\tablewidth{0pt}
\tablehead{
\colhead{Model}   &
\colhead{$h_S$ ($\degr$)}  &
\colhead{$B_{S0}$}   &
\colhead{$h_C$ ($\degr$)}  &
\colhead{$B_{C0}$}  &
\colhead{$\chi^2$} 
}
\startdata
SH only	&	$6^{pegged}$	&	$1^{pegged}$	&	\ldots	&	$\equiv0$	&	2766.2	\\
SH only	&	$0.20 \pm 0.02$	&	$1^{pegged}$	&	\ldots	&	$\equiv0$	&	422.0  \\
CB only	&	\ldots	&	$\equiv0$	&	$0.80 \pm 0.07$	&	$0.54\pm0.02$	&	339.1	\\
SH and CB	&	$6^{pegged}$	&	$1^{pegged}$	&	$0.63 \pm 0.08$	&	$0.43\pm0.02$	&	332.2	\\

\enddata
    
\end{deluxetable}

\begin{deluxetable}{lllllllll}
\tabletypesize{\scriptsize}
\tablecaption{Hapke Surge Parameters for Ten Objects 
\label{tbl6}}
\tablewidth{0pt}
\tablehead{
\colhead{Object}   &
\colhead{Band}   &
\colhead{Points}   &
\colhead{$h_S$ ($\degr$)}  &
\colhead{$B_{S0}$}   &
\colhead{$h_C$ ($\degr$)}  &
\colhead{$B_{C0}$}  &
\colhead{$N_{dof}$}  &
\colhead{$\chi^2$} 
}
\startdata
Nereid	&	V	&	196	&	$6^{pegged}$	&	$1^{pegged}$	&	$0.63 \pm 0.08$	&	$0.43 \pm 0.02$	&	194	&	332.2	\\
Huya	&	R	&	37	&	$6^{pegged}$	&	$1^{pegged}$	&	$2.3 \pm 0.7$	&	$0.47 \pm 0.12$	&	35	&	38.5	\\
Bienor	&	BVI	&	11	&	$6^{pegged}$	&	$1^{pegged}$	&	$2.5 \pm 0.4$	&	$0.69 \pm 0.06$	&	9	&	21.7	\\
Thereus	&	BVI	&	12	&	$6^{pegged}$	&	$1^{pegged}$	&	$3^{pegged}$	&	$0.43 \pm 0.04$	&	10	&	15.4	\\
1999 DE9	&	I	&	10	&	$6^{pegged}$	&	$0.4 \pm 1$	&	$1.5 \pm 0.6$	&	$0.38 \pm 0.12$	&	7	&	6.7	\\
Typhon	&	I	&	9	&	$6^{pegged}$	&	$1^{pegged}$	&	$3^{pegged}$	&	$0.7 \pm 0.2$	&	8	&	11.7	\\
Varuna	&	B	&	11	&	$6^{pegged}$	&	$1^{pegged}$	&	$1.9 \pm 0.3$	&	$1^{pegged}$	&	10	&	7.0	\\
47171	&	I	&	12	&	$6^{pegged}$	&	$1^{pegged}$	&	$2.1 \pm 0.8$	&	$1^{pegged}$	&	11	&	11.7	\\
47932	&	I	&	11	&	$6^{pegged}$	&	$1^{pegged}$	&	$1.5 \pm 0.3$	&	$1^{pegged}$	&	10	&	13.6	\\
Quaoar	&	I	&	12	&	$6^{pegged}$	&	$1^{pegged}$	&	$1.5 \pm 0.2$	&	$1^{pegged}$	&	11	&	13.0	\\

\enddata
    
\end{deluxetable}

\clearpage

\begin{figure}
\epsscale{.80}
\plotone{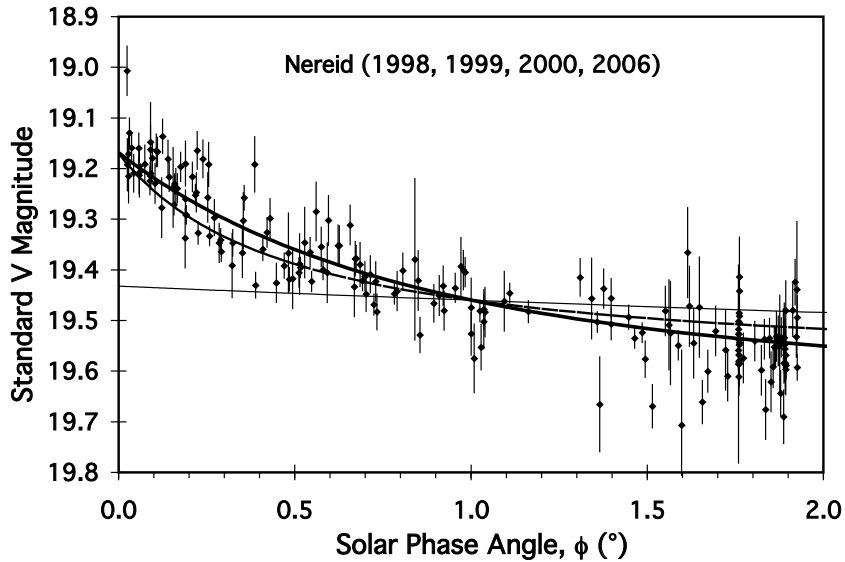}
\caption{
Nereid's phase curve fitted to SH and CB models.  For years in which Nereid is not varying greatly in brightness, its V-band phase curve is stable and well-measured.  This provides a good opportunity to experimentally test whether the shape of Nereid's phase curve corresponds to that for SH or CB, with this test being our third criterion.  The thick curve that well-fits the observations is the best fit for a CB-only model (with a chi-square of 339.1).  The best fit for the SH model with a physical constraint on the width parameter ($h_S\geq6\degr$) results in a very shallow surge, as represented by the thin nearly-flat line.  Clearly, the SH model cannot account for Nereid's phase curve.  If we allow the SH surge width to get smaller than is physically possible ($h_S=0.20\degr$), the best fit (as represented by the curve with intermediate thickness) is still bad with a chi-square of 422.0 (that is, higher than the CB-only model by 82.9).  With this, we can be confident that the observed opposition surge on Nereid is dominated by the CB mechanism.}

\end{figure}

\clearpage

\begin{figure}
\epsscale{0.5}
\plotone{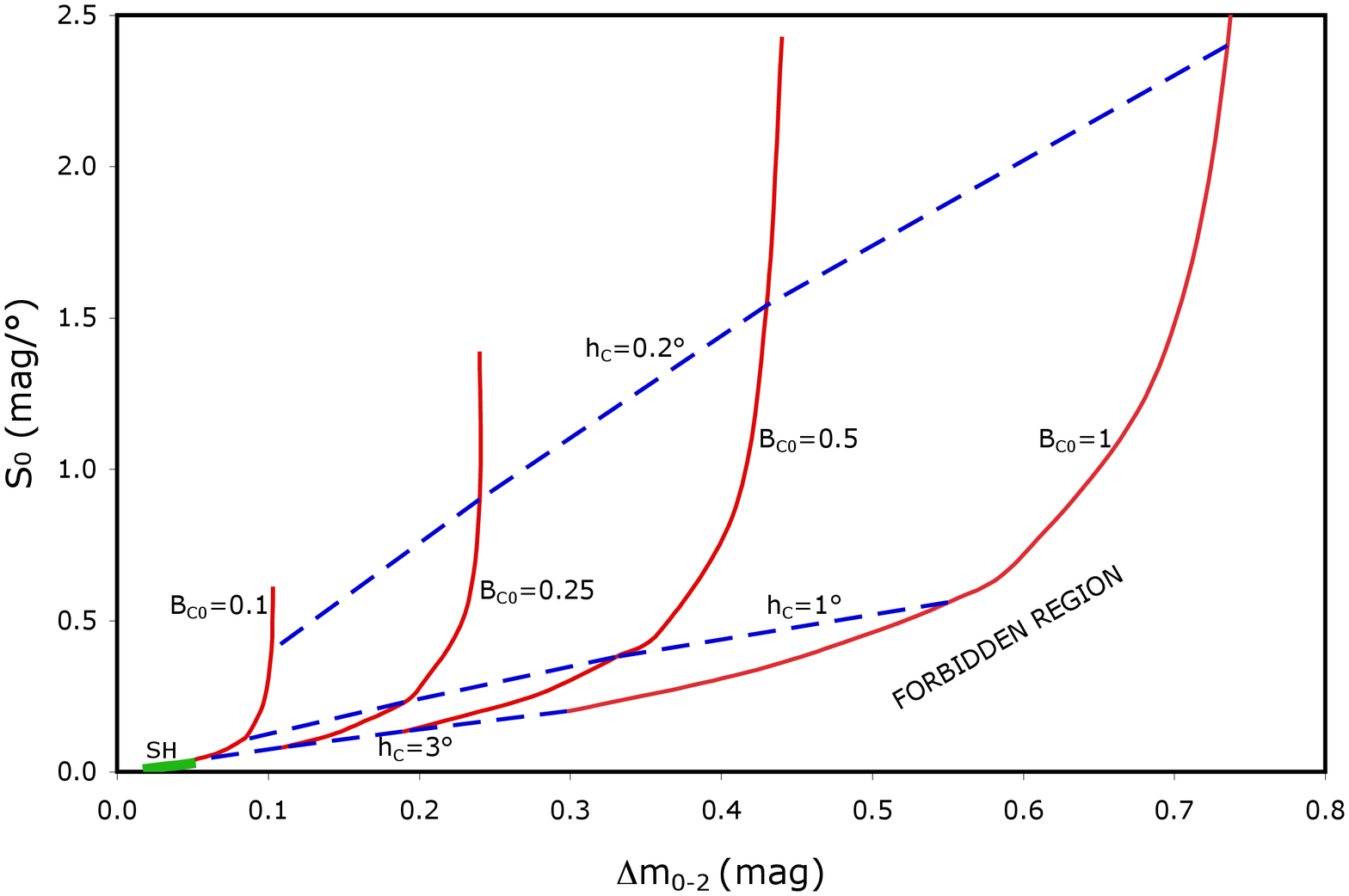}
\plotone{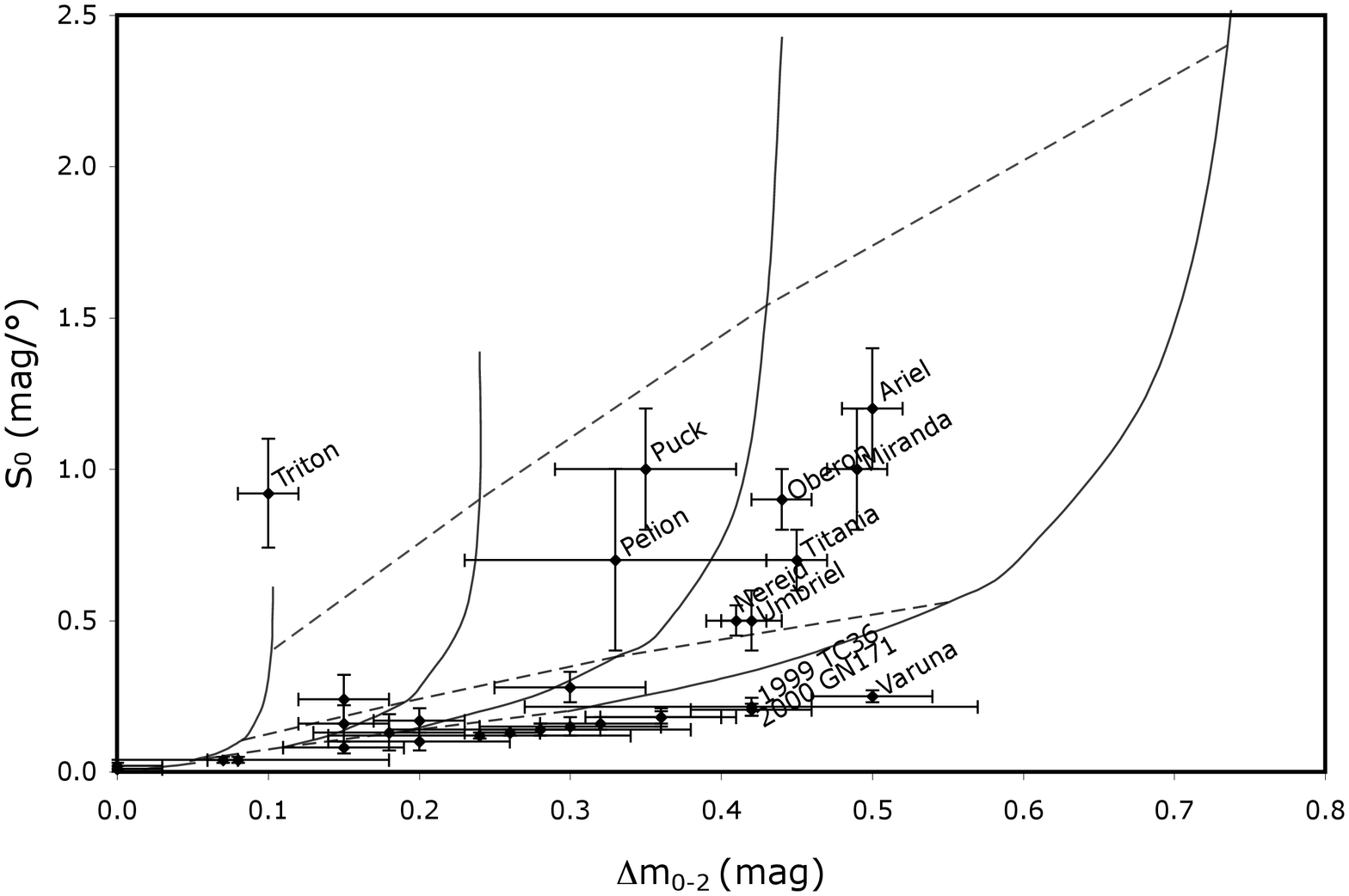}
\caption{
Observational and theoretical comparison of two surge parameters.  (a) The first panel shows the idealized curves for Hapke surge shapes (from equations 1-4).  The four smooth curves are for surfaces with $B_{C0}$   values of 0.1, 0.25, 0.5, and 1.0 for the CB mechanism and varying $h_C$.  The three dashed curves are for surfaces with $h_C$ values of $0.3\degr$, $1\degr$, and $3\degr$ for the CB mechanism and varying $B_{C0}$.  The SH mechanism is represented with a short thick curve close to the origin, with this representing surfaces with maximal SH effect.  From this plot, we see that a region in the lower right is a forbidden region where no surge can reproduce and we see that the SH mechanism is always much too small to account for almost all observed surges.  (b)  The second panel plots the individual icy bodies for which we have both $S_0$ and $\Delta m_{0-2}$.  The bodies away from the main line are labeled.  The curves from the first panel are also overplotted, and these can be used to read off the Hapke parameters for the body without a formal fit to the phase curve.  Most of the bodies fall close to the line where $S_0=0.5\Delta m_{0-2}$, which corresponds to a perfectly linear phase curve.  Positions above this line correspond to phase functions that curve so as to steepen towards low phase angles.  These positions imply that the CB mechanism have near-maximal widths, with $h_C \sim 3\degr$.  Several objects appear on the edge of this forbidden region, but formal chi-square fits (see Table 5) show that the discrepancies are not significant.}

\end{figure}

\clearpage

\begin{figure}
\epsscale{0.8}
\plotone{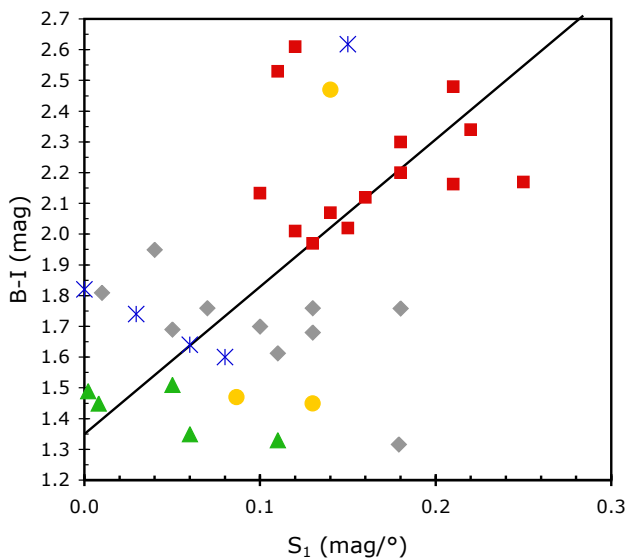}
\caption{
The surge-color correlation.  The 37 points (for all the non-moons) follow a rough correlation where the redder the object the higher the surge slope.  This correlation is highly significant (r=0.51 for a chance probability of 0.0012), although we can see that the scatter is large.  The line shows the bisector slope for this correlation.  The $S_0$ values for Eris and Sedna were used for the slope as these bodies are not observable at a phase of $1\degr$.  The symbols are squares for the Small/Red bodies, diamonds for the Small/Gray bodies, circles for the Intermediate bodies, asterisks for the Large bodies, and triangles for the Collisional bodies.}

\end{figure}

\clearpage

\begin{figure}
\epsscale{0.8}
\plotone{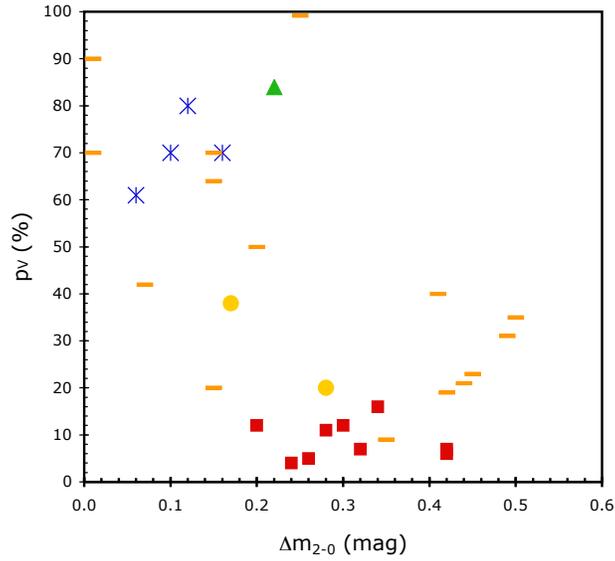}
\caption{
The surge-albedo correlation.  This plot shows the size of the surge ($\Delta m_{2-0}$) versus the albedo ($p_V$) for our 34 icy objects.  The Small/Gray bodies are not included as their surfaces have little if any ices.  The point from this plot is that all bodies with high albedo have relatively low surges, or alternatively that all bodies with large surges do not have high albedo.  Another way of saying this is that icy surfaces do not have both low surges and low albedos.  The symbols are squares for the Small/Red bodies, circles for the Intermediate bodies, asterisks for the Large bodies, triangles for the Collisional bodies, and dashes for Moons.}

\end{figure}

\clearpage

\begin{figure}
\epsscale{0.8}
\plotone{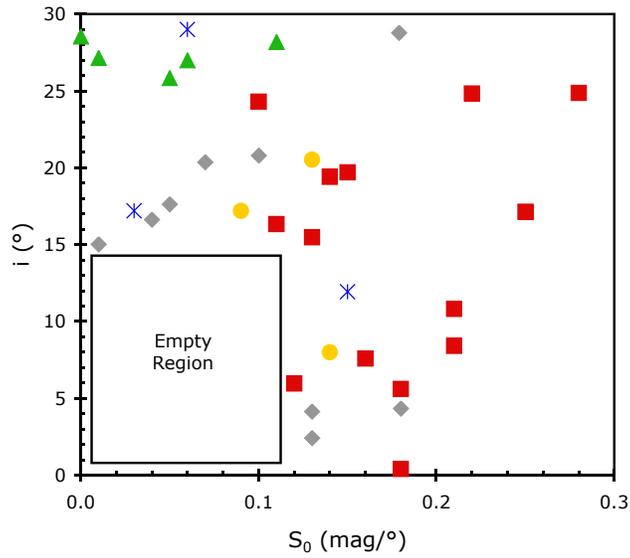}
\caption{
The surge-inclination correlation.  This plot shows the inclination ($i$) versus the surge slope ($S_0$) for our 37 non-moon objects, with one body (Eris) appearing with low surge slope off the top of the plot and two bodies (Triton and Pelion) appearing with low inclination off the right side of the plot.  No icy body has $i<15\degr$ and $S_0 < 0.12$ mag deg$^{-1}$.  The existence of this empty region is highly significant.  The symbols are squares for the Small/Red bodies, diamonds for the Small/Gray bodies, circles for the Intermediate bodies, asterisks for the Large bodies, and triangles for the Collisional bodies.}

\end{figure}

\end{document}